\documentclass[useAMS,usenatbib,fleqn]{mn2e}
\pdfoutput=1
\usepackage{graphicx}
\usepackage{natbib}
\usepackage{comment}
\usepackage{amsmath}
\usepackage{amssymb}
\usepackage{color}
\title[SPLASH: The Southern Parkes Large-Area Survey in Hydroxyl]{SPLASH: The Southern Parkes Large-Area Survey in Hydroxyl -- First Science from the Pilot Region}
\author[J. R. Dawson, A. J. Walsh, P. A. Jones et al.]
{J. R. Dawson$^{1,4}$\thanks{Email: joanne.dawson@csiro.au}, A. J. Walsh$^{2}$, P. A. Jones$^{3}$, S. L. Breen$^{1}$, M. R. Cunningham$^{3}$, \newauthor V. Lowe$^{3,1}$, C. Jones$^{4,1}$, C. Purcell$^{5}$, J. L. Caswell$^{1}$, E. Carretti$^{1}$, \newauthor N. M. McClure-Griffiths$^{1}$, S. P. Ellingsen$^{4}$, J. A. Green$^{6,1}$, J. F. G\'omez$^{7}$, \newauthor V. Krishnan$^{4,1}$, J. M. Dickey$^{4}$, H. Imai$^{8}$, S. J. Gibson$^{9}$, P. Hennebelle$^{10}$, \newauthor N. Lo$^{11}$, T. Hayakawa$^{12}$, Y. Fukui$^{12}$ and A. Mizuno$^{13}$\\
$^{1}${Australia Telescope National Facility, CSIRO Astronomy and Space Science, PO Box 76, Epping, NSW 1710, Australia}\\
$^{2}${International Centre for Radio Astronomy Research, Curtin University, GPO Box U1987, Perth, WA 6845, Australia}\\
$^{3}${School of Physics, University of New South Wales, Sydney, NSW 2052, Australia}\\
$^{4}${School of Mathematics and Physics, University of Tasmania, Private Bag 37, Hobart, TAS 7000, Australia}\\
$^{5}${Sydney Institute for Astronomy (SiFA), School of Physics, University of Sydney, NSW 2006, Australia}\\
$^{6}${SKA Organisation, Jodrell Bank Observatory, Lower Withington, Macclesfield, Cheshire SK11 9DL, UK}\\
$^{7}${Instituto de Astrof\'{\i}sica de Andaluc\'ia (CSIC), Apartado 3004, E-18080 Granada, Spain}\\
$^{8}${Department of Physics and Astronomy, Graduate School of Science and Engineering, Kagoshima University, 1-21-35 Korimoto,}\\
{Kagoshima 890-0065, Japan}\\
$^{9}${Department of Physics and Astronomy, Western Kentucky University, Bowling Green, KY 42101, USA}\\
$^{10}${Laboratoire de Radioastronomie, \'Ecole Normale Sup\'eriure and Observatoire de Paris, UMR CNRS 8112. 24 rue Lhomond}\\
{F-75231, Paris Cedex 05, France}\\
$^{11}${Departamento de Astronom\'{\i}a, Universidad de Chile, Camino El Observatorio 1515 Las Condes, Santiago, Chile}\\
$^{12}${Department of Physics and Astrophysics, Nagoya University, Chikusa-ku, Nagoya, 464-8601, Japan}\\
$^{13}${Solar-terrestrial Environment Laboratory, Nagoya University, Chikusa-ku, Nagoya, 464-8601, Japan}
}

\begin{document}

\date{Accepted 2014 Month 00. Received 2014 Month ??; in original form 2013 Month 00}

\pagerange{\pageref{firstpage}--\pageref{lastpage}} \pubyear{2014}

\maketitle

\label{firstpage}

\begin{abstract}
SPLASH (the Southern Parkes Large-Area Survey in Hydroxyl) is a sensitive, unbiased and fully-sampled survey of the Southern Galactic Plane and Galactic Centre in all four ground-state transitions of the hydroxyl (OH) radical. The survey provides a deep census of 1612-, 1665-, 1667- and 1720-MHz OH absorption and emission from the Galactic ISM, and is also an unbiased search for maser sources in these transitions. We present here first results from the SPLASH pilot region, which covers Galactic longitudes $334^{\circ}$ to $344^{\circ}$ and latitudes $\pm2^{\circ}$. Diffuse OH is widely detected in all four transitions, with optical depths that are always small (averaged over the Parkes beam), and with departures from LTE common even in the 1665- and 1667-MHz main lines. To a $3\sigma$ sensitivity of $\sim30$ mK, we find no evidence of OH envelopes extending beyond the CO-bright regions of molecular cloud complexes, and conclude that the similarity of the OH excitation temperature and the level of the continuum background is at least partly responsible for this. We detect masers and maser candidates in all four transitions, approximately 50 per cent of which are new detections. This implies that SPLASH will produce a substantial increase in the known population of ground-state OH masers in the Southern Galactic Plane.
\end{abstract}

\begin{keywords}
Galaxy: disc, ISM: molecules, masers, radio lines: ISM, surveys
\end{keywords}

\section{Introduction}

The hydroxyl radical, OH, was the first molecule detected at radio frequencies in the interstellar medium \citep[ISM;][]{weinreb63}, and is a versatile probe of ISM physics and chemistry. It exists widely throughout the Galactic ISM \citep[e.g.][]{goss68,caswell75,turner79,boyce94}, 
in local molecular clouds \citep[e.g.][]{sancisi74,wouterloot85,harju00}, high-latitude translucent and cirrus clouds \citep{grossman90,barriault10,cotten12}, associated with atomic H{\sc i} absorption features \citep{dickey81,liszt96,li03}, and extending outside the CO-bright regions of molecular clouds \citep{wannier93,allen12}. OH is an important component in diffuse gas chemistry \citep[e.g.][]{vandishoek88} and a necessary precursor to CO formation in diffuse regions \citep{black77}, where it may exhibit enhanced abundances relative to H$_2$ \citep{liszt07}. The 18 cm ground-state transitions (at 1612, 1665, 1667 and 1720\,MHz) in which OH is most commonly observed are seen both in absorption and emission, 
enabling direct measurements of the line optical depths and excitation temperatures 
\citep[e.g.][]{rieu76,dickey81,colgan89}. Relations between the four lines provide constraints on the physical properties of the gas, as well as the excitation mechanisms by which the levels are populated \citep{guibert78}. 

The first astronomical maser was also discovered in hydroxyl \citep{weaver65}. Strong OH masers occur in a wide range of astrophysical environments, including supernova remnants \citep[e.g.][]{wardle02}, evolved stars \citep[e.g.][]{sevenster97}, and high-mass star-forming regions \citep[e.g.][]{caswell87,caswell98,breen10b}, where -- compared to other masers such as methanol and water -- they are preferentially associated with later stages of the high-mass star formation process \citep[e.g.][]{forster89,caswell97}. 
Around the circumstellar envelope of evolved and dying stars, OH is preferentially associated with a later stage of stellar mass loss
compared with other species such as SiO and H$_2$O 
\citep[e.g.][]{likkel89,lintel91,nakashima03}. With a strong Zeeman splitting factor, OH masers have also been used to measure the total in-situ magnetic field, making them a unique tool for Galactic magnetic field studies \citep[e.g.][]{reid90,fish03,green12}. They also provide information on Galactic structure and dynamics complementary to that shown by the interstellar gas.

SPLASH (the Southern Parkes Large-Area Survey in Hydroxyl) is a sensitive and unbiased survey of the Southern Galactic Plane and Galactic Centre in all four ground-state transitions of OH with the Parkes 64-m telescope. 
The first phase of the survey will map the region $332^{\circ} < l < 8^{\circ}$ in Galactic longitude and $|b| < 2^{\circ}$ in Galactic latitude, at the high sensitivities necessary to detect widespread diffuse OH, as well as new, low-flux-density maser sources. The survey objectives and some relevant background on the OH ground-state transitions are described below.

\subsection{Ground state OH excitation and note on terminology}
\label{ohbasics}

\begin{figure}\includegraphics[scale=0.75]{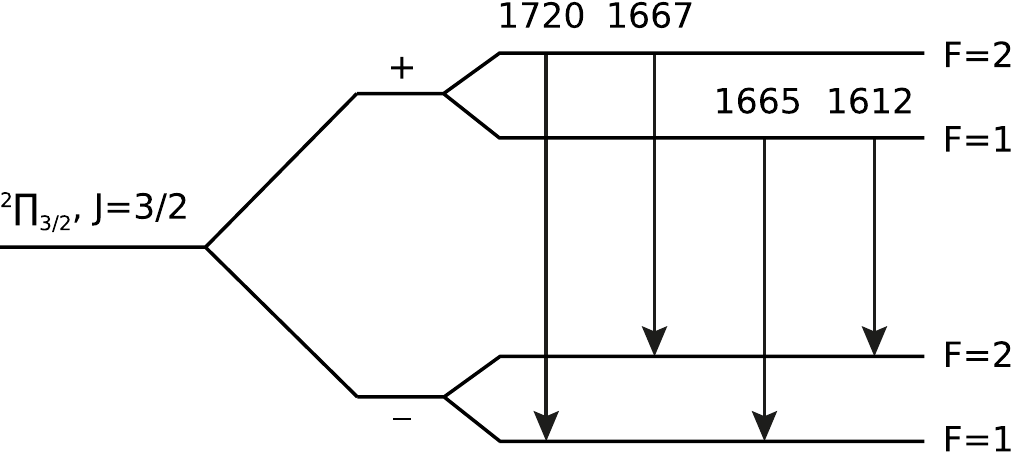}
\caption{Energy level diagram of the ground state hyperfine structure transitions of OH. Here $F$ is the total angular momentum quantum number, including the nuclear spin. Frequencies are in MHz.}
\label{ohlevels}
\end{figure}

The $^2\Pi_{3/2}, J$=3/2 ground state of OH 
is split by two processes. The first is lambda doubling -- the interaction between the rotation of the molecule and the orbital motion of the unpaired electron -- and the second is the 
hyperfine interaction between the nuclear and electronic spins. This gives rise to four sub-levels, as shown in Fig. \ref{ohlevels}, with transitions at rest frequencies of 1612.231, 1665.402, 1667.359 and 1720.530~MHz. The relative intensities of the four lines (as set by the square of the matrix element of the electric dipole moment) are 1:5:9:1 for the 1612-, 1665-, 1667-, and 1720-MHz lines respectively \citep[see][]{townes55}. In the case of local thermodynamic equilibrium (LTE), 
the optical depths of the four lines, $\tau_\nu$, 
are (to a good approximation) related by the same ratios. 

In reality, departures from LTE are common in interstellar OH. 
The satellite lines at 1612 and 1720~MHz in particular are readily inverted, and can be strongly non-thermal while the 1665- and 1667-MHz main lines remain in approximate LTE. Such behaviour arises from the structure and selection rules of the OH rotational ladder, 
which readily allow molecules to be transferred between levels with differing values of the total angular momentum quantum number, $F$, while leaving the overall population of each half of the 
$\Lambda$ doublet approximately constant \citep[see][for details]{elitzur92}. Diffuse OH line profiles commonly show an approximately symmetrical pattern of emission and absorption in the satellite lines, with excitation temperatures, $T_{\mathrm{ex}}$, that are strongly sub-thermal in one line, and either very high or negative in the other. When $T_{\mathrm{ex}}$ is negative the line is technically masing, although for diffuse OH the maser gain is very low. 


In this work we use the term `diffuse OH' to refer to the extended molecular ISM, in which maser action is either absent or very weak. The typical diagnostic signature is broad, weak line profiles with an almost symmetrical pattern of emission/absorption in the satellite lines, and main lines that are either both in emission or both in absorption. We use the term `masers' to refer to unresolved, narrow and strong emission features whose lack of matching or mirrored profiles in the full set of four lines indicates strong, high-gain maser action from a localised parcel (or parcels) of gas with strong velocity coherence.  


\subsection{Survey objectives}

\subsubsection{Diffuse OH}

Previous large-scale surveys of diffuse OH \citep[e.g.][]{robinson71,caswell74,haynes77,turner79,boyce94} have been undersampled, limited in latitude coverage, and often carried out in only one or two lines (see \citealp{turner79} for a comprehensive tabulation of all surveys prior to that work). Crucially, they have also lacked the sensitivity to detect anything but the strongest diffuse OH emission and absorption. 

SPLASH is an order of magnitude more sensitive than these past surveys, and provides a high-sensitivity, high-velocity-resolution, 
fully-sampled and wide-area census of diffuse OH in the inner Galaxy. 
A major outcome of SPLASH will be the ability to 
quantify key parameters of the diffuse OH distribution on Galactic scales, such as its scale height and degree of concentration in spiral arms, as well as physical properties such as masses, column densities, optical depths and excitation states. The ability to trace 
patterns of satellite-line excitation across large sections of the Galaxy provides a promising new tool with which to probe the density and ambient IR radiation field \citep[e.g.][]{guibert78}, and may also provide a means of disentangling physically distinct structures that are blended in the spatio-velocity domain. Another key aim is to assess the effectiveness of OH as a probe of low-extinction, partially-molecular material in which CO abundances are low \citep[e.g.][]{wannier93,liszt96,allen12,cotten12}. This will be carried out via comparisons with existing CO \citep{mizuno04} and H{\sc i} \citep{mcclure05,kalberla10} surveys. 

An important corollary aim of SPLASH is to provide vital short-spacing data for the upcoming GASKAP survey, which will use the Australia Square Kilometre Array Pathfinder (ASKAP) telescope to image H{\sc i}, diffuse OH and OH masers in the 1612-, 1665- and 1667-MHz lines throughout the Galactic Plane, Magellanic Clouds and Magellanic Stream \citep{dickey13}. 





\subsubsection{Maser science}

SPLASH is a fully-sampled blind survey for all four ground-state maser transitions with sensitivities and coverages that improve (to varying degrees) on previous work 
\citep[e.g.][]{caswell83,caswell87,sevenster97,caswell98,sevenster01}. The primary aim with regards to masers is to provide a sensitive and unbiased census of 1612-, 1665-, 1667- and 1720-MHz OH masers in the Fourth Quadrant of the Galactic Plane and the Galactic Centre region, which will form the basis of a wide range of follow-up studies. 
While SPLASH observations themselves are in total intensity, follow-up high-resolution observations towards newly detected sources in full Stokes are underway at the Australia Telescope Compact Array (ATCA).

OH main-line masers arise primarily in high-mass star forming regions. Science targets for the maser portion of the survey and its associated follow-up work include consolidation of the role of OH 1665- and 1667-MHz masers as evolutionary signposts of the high-mass star formation process 
\citep[e.g.][]{ellingsen07, breen10a}, as well as the identification of new low flux density objects to supplement the sources in the `MAGMO' project, which aims to trace Galactic magnetic fields using ground state maser Zeeman splitting measurements 
\citep{green12}. 
1612-MHz masers primarily arise from evolved stars, with a smaller fraction arising from star forming regions. The detection of new 1612-MHz sources will facilitate large-scale statistical comparisons with SiO and H$_2$O masers in evolved stars from the Asymptotic Giant Branch (AGB) to planetary nebula (PN) phases, with the aim of clarifying the evolutionary sequence of observed maser transitions in these objects \citep[e.g.][]{lewis89,nyman98,gomez07}. SPLASH will also detect new examples of the rarer 1720-MHz masers, which are most commonly associated with star forming regions \citep[e.g.][]{caswell99,caswell04}, often accompanied by main-line masers; a smaller, quite distinct, class is associated with supernova remnants \citep[e.g.][]{green97}, where the 1720-MHz masers are not accompanied by any other transition. 

\section{Observations and data processing}
\label{obs}

\subsection{Observations}
The pilot region covers the range $334^{\circ} < l < 344^{\circ}$ in Galactic longitude and $|b| < 2^{\circ}$ in Galactic latitude.
Observations were made between 2012 May and November with the Australia Telescope National Facility (ATNF) Parkes 64-m telescope. The receiver was the H-OH receiver, with a standard feed that provided two orthogonal linear polarisations. Data were taken in on-the-fly (OTF) mapping mode, in which spectra are taken continuously as the telescope is scanned across the sky. The survey region was divided into $2\times2$ degree tiles, each of which was mapped a total of ten times to achieve target sensitivity. Repeat maps were scanned alternately in the Galactic latitude and longitude directions to minimise scanning artefacts. The scan rate was $34$ arcsec~s$^{-1}$, with data output every 4~s at intervals of $2.3$ arcmin, and the spacing between scan rows was $4.2$ arcmin. This oversamples the beam (which has a FWHM of $12.6$ arcmin at 1720~MHz) both perpendicular and parallel to the scan direction. Off-source reference spectra were taken every two scan rows, where the off-source position for each map was chosen to minimise the elevation difference between the reference position and the map throughout the course of observations. All reference positions were observed for a total integration time of 20 minutes prior to the main survey to ensure no emission or absorption was present. 
The ATNF standard calibrator source PKS B1934-638 was observed once per day in two orthogonal scans. 

The system temperature was monitored using a rapidly switched noise source injected at the front end of the receiver, and was recorded every integration cycle. Typical system temperatures were $25$--$30$ K, excluding pointings towards bright continuum sources within the Galactic Plane. Data were recorded simultaneously in three frequency bands centred on 1612, 1666 and 1720~MHz, each with 8192 spectral channels over a bandwidth of 8~MHz. This corresponds to a velocity coverage and channel width of $\sim1400$~km~s$^{-1}$ and 0.18~km~s$^{-1}$ respectively. 

\subsection{Data reduction}
\label{reduction}

Bandpass calibration, flux density calibration and velocity scaling was carried out with the {\sc ASAP}\footnote{http://www.atnf.csiro.au/computing/software/asap/refman/} and {\sc Livedata}\footnote{http://www.atnf.csiro.au/computing/software/livedata/index.html} 
data reduction packages, in combination with {\sc python} modules currently under development as part of the SPLASH data reduction pipeline. Bandpass calibration was performed using off-source reference spectra according to:
\begin{equation}
S^*=\frac{P_{\mathrm{ON}}-P_{\mathrm{OFF}}}{P_{\mathrm{OFF}}}~T_{sys,\mathrm{OFF}}
\end{equation}
where $S^*$ is the flux density of the source position, $P_{\mathrm{ON}}$ and $P_{\mathrm{OFF}}$ are the on-source and off-source power measured by the telescope, and $T_{sys,\mathrm{OFF}}$ is the system temperature at the off-source position (in Jy). 
Residual spectral baseline structure was removed by 
flagging bright spectral lines, 
performing a linear interpolation over the missing channels, then heavily smoothing the remaining spectrum with a Gaussian kernel of $\sigma=220$ channels ($\sim40$~km~s$^{-1}$) to produce a model of the continuum flux and residual baseline shape. This model is then subtracted from the original spectrum to produce continuum-subtracted, baseline-corrected spectral line data. This process was carried out on each raw spectrum (4 s integration time) prior to gridding, and later again on the gridded cube, for which the better signal-to-noise resulted in greatly improved baseline solutions.

More precise calibration factors were derived for each frequency and polarisation from the daily observations of PKS B1934-638, assuming standard (unpolarised) flux densities of 14.34, 14.16 and 13.98 Jy at 1612, 1666 and 1720~MHz \citep{reynolds94}. Small correction factors of between 1.1 and 1.3 were derived, which 
remained stable to within $\sim1$ per cent over the full period of the observations. The conversion from flux density to a main-beam brightness temperature scale ($T_{\mathrm{mb}}$) was then performed assuming standard main-beam gains for the H-OH receiver of 1.30, 1.34 and 1.39 Jy K$^{-1}$ for 1612, 1666 and 1720~MHz\footnote{http://www.parkes.atnf.csiro.au/cgi-bin/public\_wiki/wiki.pl?H-OH}. From hereon any reference to observed brightness temperature is referring to this main-beam temperature.

Gridding of the spectra into spatio-velocity cubes was performed with the {\sc Gridzilla}$^2$ software package \citep{barnes01}. We use a Gaussian smoothing kernel of $20\arcmin$ truncated at a cutoff radius of $10\arcmin$, which results in an effective resolution (FWHM) of $\sim15.5$ arcmin. This relatively large kernel size is chosen to improve the sensitivity to extended sources and optimise the spectral line cubes for weak, extended signal. The statistic used to compute the brightness temperature in a given pixel is a weighted mean with censoring of outliers, chosen for its robustness against RFI. The two orthogonal polarisations are summed during gridding to produce total intensity (Stokes I) cubes. 
A separate set of Jy-scaled, point-source-optimised cubes were created for maser science applications, by enabling the `beam normalisation' option in {\sc Gridzilla}. Further details of the gridding algorithm can be found in \citet{barnes01} and the {\sc Gridzilla} in-package documentation.

Post-gridding, low-level signal was discovered in two reference positions belonging to both the positive and negative latitude maps at $336^{\circ} < l < 338^{\circ}$. This was corrected to the first order by removing the affected velocity range ($-5.5 < v_{LSR} < -2.5$~km~s$^{-1}$) from the raw reference spectra and linearly interpolating over the missing channels. The residual effect on the data is small, and has no significant effect on the analysis in this work. Accurate fitting and subtraction of reference position signal will be carried out in the final processing pipeline.

The final rms sensitivity to extended emission in the gridded cubes is between $\sim25$--$70$ mK in a 0.18~km~s$^{-1}$ spectral channel, with a mean of $\sim35$ mK, where the highest noise is seen at positions coincident with strong continuum sources. 
The mean rms point-source sensitivity in the maser-optimised cubes is $\sim65$ mJy in a 0.18~km~s$^{-1}$ channel. For the purposes of much of the diffuse OH work, spectra are binned by 4-channels in the velocity dimension, increasing the channel width to 0.7~km~s$^{-1}$ and reducing the mean noise to $\sim16$ mK. Unless otherwise stated, all analysis of diffuse OH in this work makes use of these binned cubes. Further improvements, particularly in baseline quality and residual RFI mitigation, are to be implemented in the final processing pipeline. 
 
\subsection{Continuum maps and system stability}

The 
model of the continuum and baseline residuals is also calibrated and gridded as above, then averaged in frequency to produce complementary continuum images. 

The measured continuum flux is sensitive to additional contributions to the incoming power such as ground radiation, as well as variations in system gain. These effects are generally elevation-dependent, and may also vary temporally on both long and short timescales. 
In SPLASH, 
off-source spectra are taken every $\sim10$ minutes, and the maximum elevation difference between the furthest corner of a map and its corresponding off-source reference position never exceeds 4 degrees (and is typically less than 2 degrees). The gain terms contained within $P_{\mathrm{ON}}$ and $P_{\mathrm{OFF}}$ in equation 1 are therefore expected to be reasonably similar, and errors arising from gain drift are not severe. In addition, skydip observations confirm a maximum error of $\sim0.4$ K in the main beam brightness temperature arising from the elevation difference between ON and OFF positions. This value is always $<10$ per cent of even the weakest continuum emission in our survey region.

This stability is quantitatively confirmed by examining repeat maps of the same region taken at multiple epochs and elevations. The per-pixel standard deviation for a given position in repeat observations of the same map is typically $\sim0.05$--0.6 K; or 
within $<10$ per cent of the mean brightness temperature for 99 per cent of pixels. Differences are not correlated with LST, UTC, absolute elevation or Julian date. 


\begin{figure*}\includegraphics[scale=0.95]{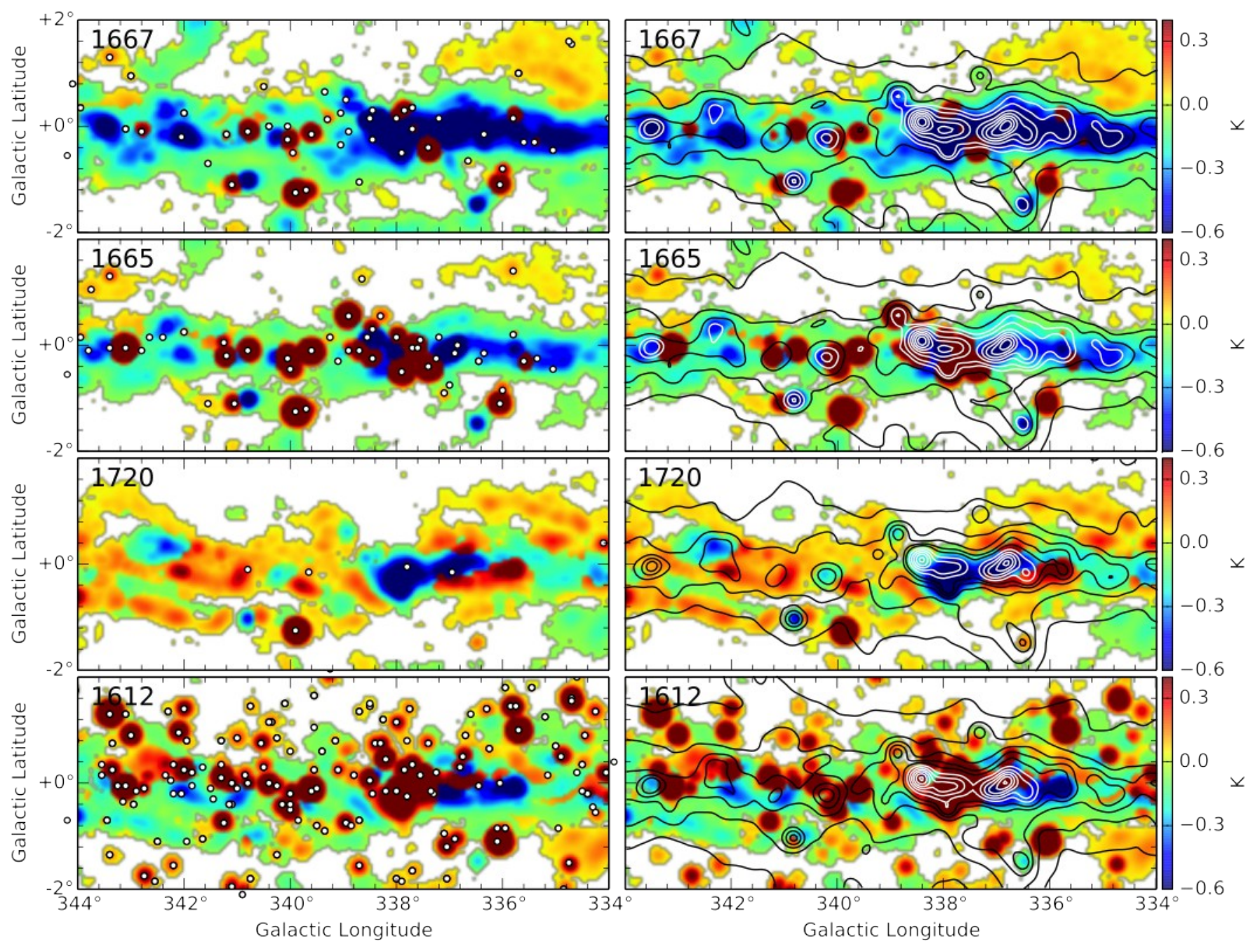}
\caption{Combined peak emission and peak absorption maps for all four ground state OH lines in the SPLASH pilot region, created by plotting the most extreme value of the brightness temperature at each spatial position. Only detections significant at the $4\sigma$ level are shown. Small black-bordered white circles overlaid on the left-hand panels show the peak positions (to the nearest $3$ arcmin pixel) of masers and maser candidates. Contours overlaid on the right-hand panels show the continuum brightness temperature at the line rest frequency. Contour levels run from 10.0 to 20.0 K at intervals of 2.5 K, then from 20.0 to 50.0 K at intervals of 5.0 K. Contour colours are chosen to aid visibility and have no physical meaning. The panels are ordered with the main lines (1667 and 1665 MHz) first and satellite lines (1720 and 1612 MHz) second to highlight the corresponding or conjugate nature of the four transitions.}
\label{mom-4tiles}
\end{figure*}

A fundamental limitation of referenced observing is that 
we may only directly measure the \textit{difference} in continuum brightness between the on-source and off-source positions. 
In order to recover the absolute value, we must therefore obtain an estimate of the brightness temperature at 1.6--1.7 GHz at the survey reference positions. For this we make use of the S-PASS 2.3 GHz \citep{carretti13} and HIPASS 1.4 GHz \citep{calabretta13} continuum surveys of the Southern sky. The HIPASS data use a full beam temperature scale, are absolutely calibrated, and include the 2.73 K cosmic microwave background (CMB). S-PASS data includes no CMB contribution and no zero-level offset, and has an uncertainty in its zero level of 0.1 K (Carretti, private communication). Both datasets have an effective beam size of $14.5$ arcmin (S-PASS was originally $10.75$ arcmin, smoothed here to the HIPASS resolution for the goals of this work). 

The SPLASH reference positions are located at $|b|>2.25^{\circ}$, where the continuum emission is primarily diffuse and extended compared to the Parkes beam. In order to compute a spectral index at each reference position, we first subtract the zero level offset of 3.3 K from the HIPASS data \citep{calabretta13}, where this value contains the 2.73 K CMB and a $\sim0.6$ K contribution from the Galactic diffuse synchrotron emission in the lowest emission areas. The map is then rescaled to a main beam temperature scale assuming a main beam efficiency of $\eta_{\mathrm{mb}}=0.60\pm0.08$. This value and its uncertainties are estimated from the efficiency implied by the scaling of \citet[][$\eta_{\mathrm{mb}}=0.52$]{calabretta13}, 
and that measured for the central beam of the Parkes multibeam receiver \citep[][$\eta_{\mathrm{mb}}=0.69$]{staveley96}. Spectral indices at each reference position were then computed from the HIPASS and S-PASS data, and used to estimate the continuum brightness temperature at 1612, 1666 and 1720~MHz respectively. A final zero level correction of $+3.1$ K was then applied, which consists of the CMB plus a $\sim0.4$ K contribution from the Galactic diffuse synchrotron emission. This was estimated from the 1.4 GHz value 
by assuming a spectral index of 2.7 \citep{platania03}. The uncertainty propagated through from the choice of main beam efficiency is $\pm0.5$ K. 

The reference position brightness temperatures are added to their corresponding continuum map tiles prior to gridding, to produce absolutely calibrated 1612, 1666 and 1720\,MHz continuum maps. The resulting maps vary smoothly across tile boundaries, confirming that the magnitude of the corrections are accurate to within a few tenths of a Kelvin. Overall, we consider the uncertainty on the final continuum maps to be $\lesssim1.0$ K in $T_{\mathrm{mb}}$. 

\section{Results}
\label{results}

OH is widely detected in all four transitions throughout the pilot region. Fig. \ref{mom-4tiles} shows combined peak emission and peak absorption maps for each line, created by plotting the most extreme value of the brightness temperature, $T_\mathrm{b}$ at each spatial position. 
Continuum contours and the peak positions of maser sources are also shown. 
Fig. \ref{ch1} shows channel maps of the same region, in which the data have been binned in the velocity dimension to a channel width of 3.6~km~s$^{-1}$ in order to improve the signal-to-noise. 

\begin{figure}\includegraphics[scale=0.75]{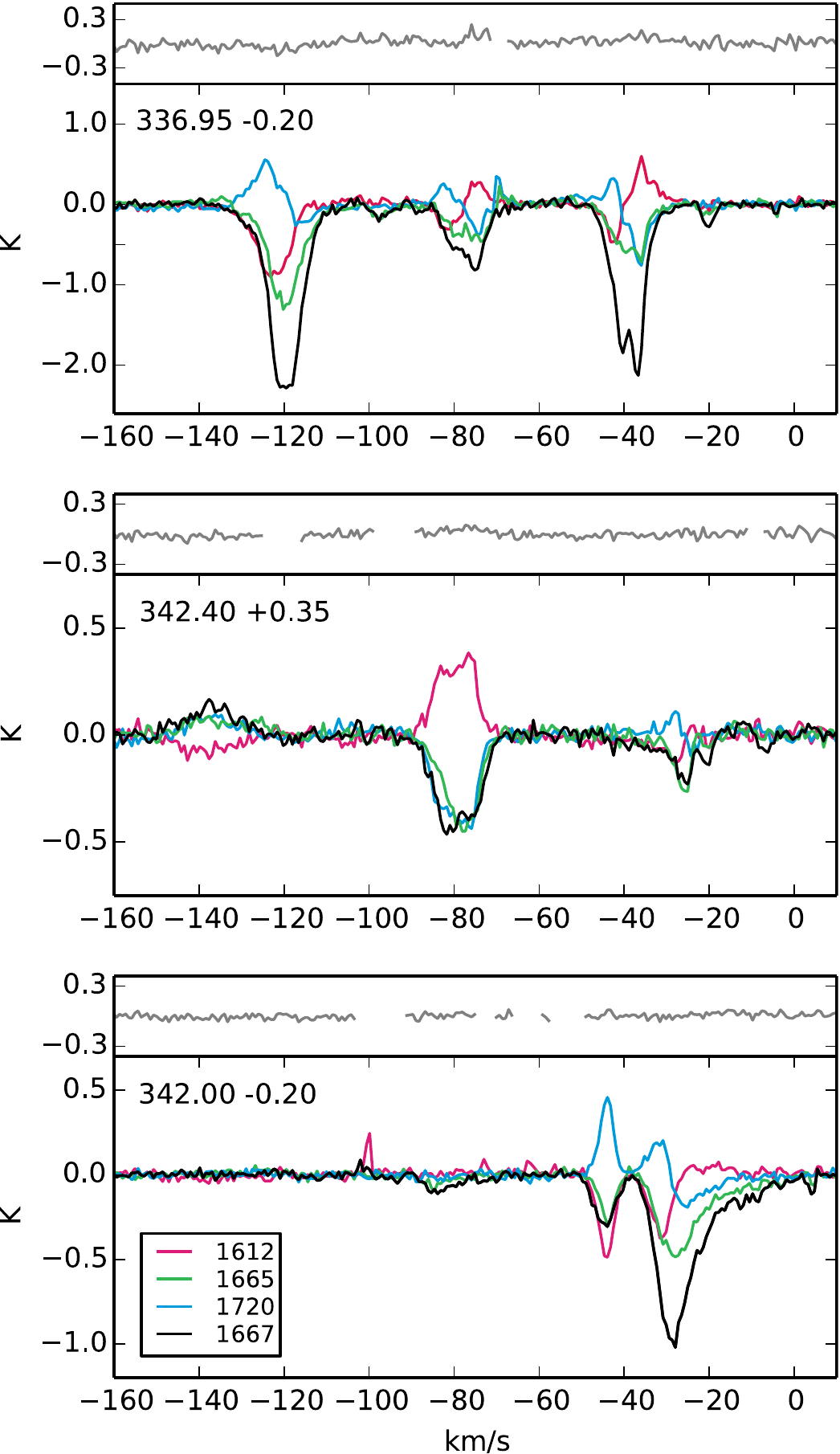}
\caption{Example spectra showing primarily diffuse OH. The grey spectra above each panel show $[T_{\mathrm{b}}(1612)+T_{\mathrm{b}}(1720)]-[T_{\mathrm{b}}(1667)/9.0 + T_{\mathrm{b}}(1665)/5.0]$, with voxels in the vicinity of maser emission flagged out (blank ranges). 
The diffuse OH lines show their characteristic pattern -- main-line signal with $|T_{\mathrm{b}}(1667)| \gtrsim |T_{\mathrm{b}}(1665)|$ (usually in absorption), with a symmetrical pattern of emission and absorption in the satellite lines. The data in these figures have been binned to a channel width of 0.7~km~s$^{-1}$.} 
\label{specs}
\end{figure}

The main lines are dominated by a band of strong absorption along the Galactic Plane, much of which arises from the Scutum-Crux Arm at $-50\lesssim v_{LSR}\lesssim-30$~km~s$^{-1}$. The most prominent main-line emission is seen at velocities more negative than $-120$~km~s$^{-1}$, consistent with a location in the 3 kpc arm \citep[see e.g.][]{dame01,green11}, and behind the bulk of the continuum-emitting material along the line of sight. Prominent absorption features at higher latitudes of $|b|\gtrsim1.0^{\circ}$ are relatively local gas ($-25\lesssim v_{LSR}\lesssim +5$~km~s$^{-1}$) seen in absorption against the diffuse Galactic background. Main-line masers lie clustered along the Galactic Plane, as expected for sources that predominently trace star-forming regions. With the exception of these bright maser sources, the 1665-MHz line is generally weaker than the 1667-MHz line, as expected from 
their differing transition strengths. Example spectra are shown in Fig. \ref{specs}.
 
The 1612- and 1720-MHz satellite lines show the expected anomalous excitation pattern, with one transition seen in emission and the other in absorption (see Figs. \ref{specs} and \ref{ch1}). The most common sense of the inversion in this region is emission in the 1720-MHz line and absorption in the 1612-MHz line; however, both senses are seen commonly throughout the cube. This is consistent with the findings of \citet{haynes77} who observed this region at considerably lower sensitivity. With the exception of masers, the satellite-line brightness temperatures are always of a similar magnitude, and where differences exist, the satellite line showing the same behaviour as as the main lines is the stronger of the pair. This phenomenon, and its interpretation, are discussed further in section \ref{sumrule}. 

The absolute brightness temperatures of the satellite lines are generally smaller than the main lines, which is the a-priori expected behaviour given their smaller transition strengths. However, this is not always the case, and the satellite lines are even detected in some places where the main lines are not. Excluding masers, we find that $\sim10$ per cent of $4\sigma$ OH detections are detected in the satellite lines alone -- a fact that highlights their largely-untapped potential as diffuse molecular gas tracers \citep[see also][]{turner82}. A prominent example is the thick filament of 1720-MHz emission (and 1612-MHz absorption) seen between $341^{\circ} < l < 343^{\circ}$, $-1.3 < b < -0.7^{\circ}$, which is completely absent at 1665 and 1667~MHz. Such behaviour is expected where the main-line excitation temperatures are close in magnitude to the background continuum level, but 
$T_{\mathrm{ex}}(1612)$ and $T_{\mathrm{ex}}(1720)$ are not. This is discussed further in section \ref{ohvsco}. 

The 1612-MHz satellite transition shows an abundance of bright masers, most of which show the double-horned profile characteristic of the expanding shells of evolved stellar sources. Unlike the main-line masers these 1612-MHz masers show only minimal clustering around the midplane. 1720-MHz masers are rare, with only six detections in this region. We return to the subject of the maser population in section \ref{masers}.

\section{Discussion}



\subsection{The brightness temperature sum rule and constraints on optical depth}
\label{sumrule}

For small optical depths, background continuum emission with a flat brightness temperature spectrum, and $|T_{\mathrm{ex}}| \gg h\nu/k ~(\approx0.08$ K), the intensities of the four ground state transitions are intrinsically related by the brightness temperature ``sum rule'' \citep[see e.g.][]{robinson67, brooks01},
\begin{equation}
\frac{T_{\mathrm{b}}(1667)}{9} + \frac{T_{\mathrm{b}}(1665)}{5} = T_{\mathrm{b}}(1612) + T_{\mathrm{b}}(1720).
\label{tbsumrule}
\end{equation}
We find that, within the uncertainties, essentially all diffuse OH detections in the SPLASH pilot region follow this relationship. Blanking strong maser emission from the cube, and restricting analysis to voxels detected at the $4\sigma$ level in at least one line, the quantity $|T_{\mathrm{b}}(1667)/9 + T_{\mathrm{b}}(1665)/5 - T_{\mathrm{b}}(1612) - T_{\mathrm{b}}(1720)|$ is only significantly greater than zero (at the $3\sigma$ level) for 0.3 per cent of the sample -- consistent with a null result. (See also Fig. \ref{specs} for example spectra.)

We construct a simple line model to examine the origins of this behaviour. Excitation temperatures are assigned to three of the lines, and the fourth determined by the excitation temperature sum rule, 
\begin{equation}
\frac{\nu_{1667}}{T_{\mathrm{ex}}(1667)} + \frac{\nu_{1665}}{T_{\mathrm{ex}}(1665)}= \frac{\nu_{1612}}{T_{\mathrm{ex}}(1612)} + \frac{\nu_{1720}}{T_{\mathrm{ex}}(1720)},
\end{equation}
\noindent which is a direct consequence of the definition of $T_{\mathrm{ex}}$ and is true under all conditions. 
A total OH column density is assumed, and divided amongst the energy levels according to the Boltzmann equation and line excitation temperatures. 
Optical depths are then calculated from the usual definition of opacity according to
\begin{equation} 
\tau_v=\frac{c^3}{8\pi\nu_0^3} \frac{g_u}{g_l} N_l A_{ul} (1-e^{-h\nu_0/k T_{\mathrm{ex}}})~\phi_v,
\end{equation}
\noindent where $\nu_0$ is the line rest frequency, $A_{ul}$ is the Einstein-A coefficient of the transition, ${g_u}$ and ${g_l}$ are the degeneracies of the upper and lower levels, $N_l$ is the column density of particles in the lower energy level, and $\phi_v$ is a normalised Gaussian profile with a velocity width of $\Delta v$ (where all quantities are in SI units). Finally, a 1667-MHz continuum background temperature and spectral index are assumed, and the line brightness temperature profiles computed from the basic expression for radiative transfer, 
\begin{equation}
T_{\mathrm{b}}(\nu)=(T_{\mathrm{ex}} - T_{\mathrm{c}})(1-e^{-\tau_v}),
\label{radtran}
\end{equation}
\noindent where this is expression describes the case where 
all the continuum emission, $T_{\mathrm{c}}$, lies behind the OH gas. We thus construct simple model profiles for a range of excitation states, column densities and continuum background temperatures, and may assess the range of conditions under which we would expect to detect deviations from equation (\ref{tbsumrule}) in the SPLASH cubes. 


We find that even for small $\tau$, some deviation is expected as a result of the frequency dependence of $T_{\mathrm{c}}$, and in some cases from anomalous excitation (when $|T_{\mathrm{ex}}|$ is unusually small). 
However, with the exception of particularly strong satellite-line inversions, the magnitude of the deviations is generally within the noise in the SPLASH datacube. Conversely, for $\tau_{1667}\gtrsim1$, the rule rarely holds, and strong deviations are readily observable for the majority of tested excitation states and continuum background temperatures. While this analysis is simplistic, 
the lack of significant departures from the rule over the entire datacube strongly suggests that the optical depths of the OH (averaged over the Parkes beam) are small. This is consistent with the available direct measurements of in-Plane OH optical depths along limited sightlines, which typically find $\tau_{1667}\lesssim0.1$. \citep{rieu76,crutcher79,dickey81,colgan89}. The present work strongly suggests that the assumption of small optical depths is valid over large sections of the inner Galactic Disk; at least on the scales probed by the Parkes beam.



\begin{figure}\includegraphics[scale=0.85]{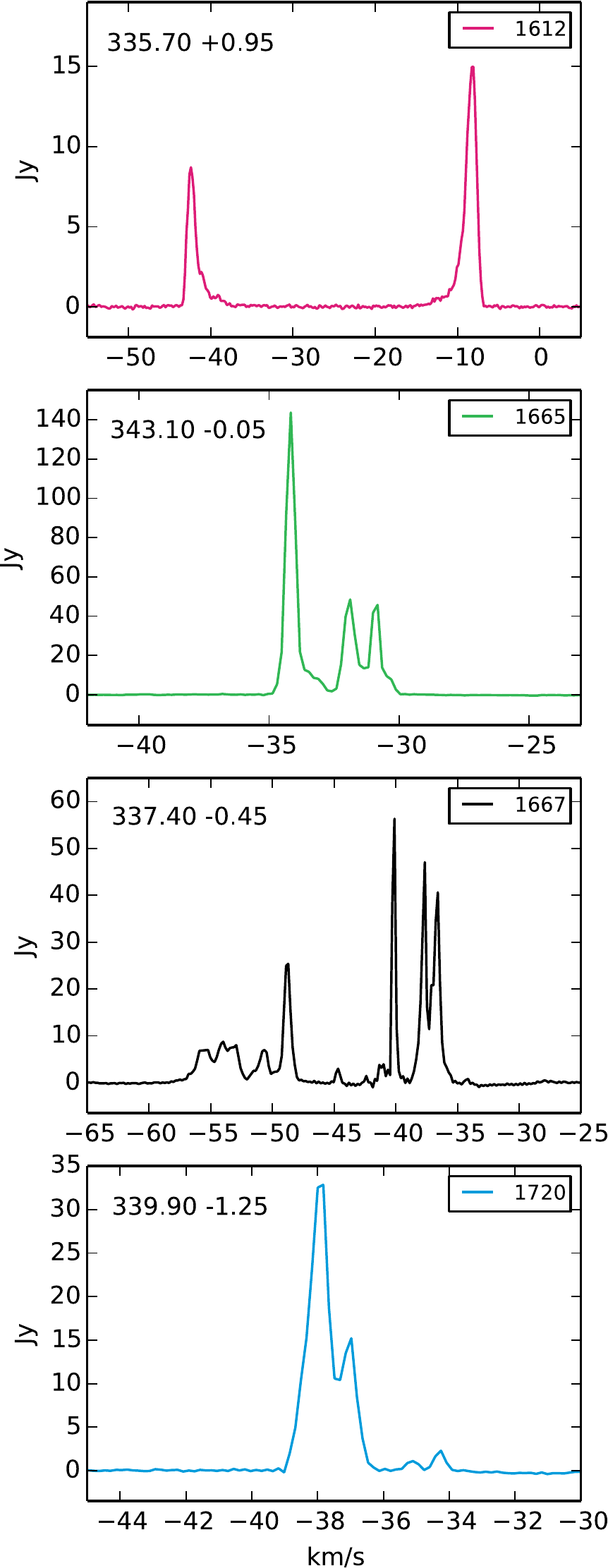}
\caption{Example spectra showing strong maser emission. The data in these figures are extracted from 
point-source-optimised cubes scaled in Janskys (see section \ref{reduction}), and have a channel width of 0.18~km~s$^{-1}$. The precise designations of these known sources from the literature are (from the top downwards): 335.717+0.991 \citep{sevenster97} -- an evolved stellar source; 343.127-0.063, 337.405-0.402 \citep{caswell98} -- main-line masers arising in high-mass star forming regions; and 339.884-1.259 \citep{caswell04} -- a highly variable 1720-MHz maser again arising from a high-mass star forming region.}
\label{maserspecs}
\end{figure}

\subsection{Evidence for widespread main-line anomalies and limitations of the LTE assumption}

The assumption that the main lines are in local thermodynamic equilibrium is sometimes used as a tool in the analysis of diffuse OH data 
\citep[e.g.][]{heiles69,knapp73,mattila79b,wouterloot84,wouterloot85,li03}. In LTE, the main-line optical depths are related by $\tau_{1667}=1.8\,\tau_{1665}$, and departures from this ratio in the line brightness temperatures are assumed to arise purely from optical depth effects (rather than excitation effects). 
This approach is typically rationalised from an observational perspective by the confirmation of main-line intensity ratios in the LTE range of $R_{1667/1665} = T_{\mathrm{b}}(1667)/T_{\mathrm{b}}(1665) = 1.0$--1.8, where these values correspond to the optically thick and optically thin limits, respectively. The theoretical rationale for approximately LTE main lines has already been discussed in \S\ref{ohbasics}.

Nevertheless, `main-line anomalies' are frequently reported in the literature -- either in the form of $R_{1667/1665}$ values outside the permitted range \citep[e.g.][]{turner73,sancisi74,kazes77,dickey81,grossman90}, or from explicit measurements of  $T_{\mathrm{ex}}(1667)$ and $T_{\mathrm{ex}}(1665)$ via emission/absorption observations against distant continuum sources \citep{crutcher77,crutcher79,colgan89}. Of course, even a `thermal' line ratio is not a conclusive argument in favour of LTE, since permitted ratios may arise for various non-LTE combinations of $T_{\mathrm{ex}}$ and $\tau$. As has been discussed in detail by \citet{crutcher79}, such anomalies can potentially cause large errors in derived properties if LTE is erroneously assumed. 

An examination of voxels detected at the $4\sigma$ level in both main lines in the SPLASH cube reveals that $R_{1667/1665}$ falls in the 1.0--1.8 range for approximately half. Once uncertainties are taken into account, only $\sim25$ per cent fall robustly within the LTE range, and $\sim10$ per cent fall robustly outside it, while the remaining $\sim65$ per cent are ambiguous at the $3\sigma$ level. The physical interpretation of these numbers must be approached with caution, since a given velocity may contain multiple overlapping components with differing densities and excitation states. However, we may use this simple measure to select voxels for which the LTE assumption produces valid mathematical solutions to the transfer equations, and examine the extent to which the assumption may be justified.

Under the assumption that all continuum emission, $T_{\mathrm{c}}$, lies behind the OH cloud, the radiative transfer equations for the main lines in LTE are given by:
\begin{equation}
\noindent T_{\mathrm{b}}(1667)=(T_{\mathrm{ex}} - T_{\mathrm{c}})(1-e^{-\tau_{1667}})
\end{equation}
\vspace{-7mm}
\begin{equation}
T_{\mathrm{b}}(1665)=(T_{\mathrm{ex}} - T_{\mathrm{c}})(1-e^{-\tau_{1667}/1.8}),
\end{equation}
\noindent and may be solved exactly for $T_{\mathrm{ex}}$ and the two optical depths. Restricting our analysis to voxels where a physical solution exists ($R_{1667/1665}$ from 1.0--1.8), and to velocities at which most OH may be reasonably assumed to lie in front of the continuum ($|v_{LSR}| < 20$~km~s$^{-1}$), we may compute LTE solutions for $T_{\mathrm{ex}}$ and $\tau$ throughout the pilot region. The results of this experiment are telling. We find that $T_{\mathrm{ex,LTE}}$ is closely locked to the background continuum temperature, with $|T_{\mathrm{c}} - T_{\mathrm{ex,LTE}}| < 1$ K. $\tau_{\mathrm{LTE}}$ fluctuates wildly, both spatially and across line profiles, frequently reaching very high values ($\sim5$--15) that are inconsistent with the results of the previous section (\S\ref{sumrule}). 


This clearly non-physical behaviour arises as a consequence of forcing an LTE-solution to main lines whose excitation temperatures are in reality not identical, a fact that was also recognised by \citet{crutcher79}. He suggests that main-line anomalies of $|T_{\mathrm{ex}}(1667) - T_{\mathrm{ex}}(1665)| \sim 0.5$--2.0 K are probably common throughout the diffuse ISM, based on analysis of a small number of sight lines. Our results reach a similar conclusion for a much larger area, and strongly suggest that anomalous excitation in the OH main lines is the norm -- at least in this part of the Galactic Disk, and more likely in the general case. 

\subsection{OH vs CO}
\label{ohvsco}

OH has been found to be an effective tracer of diffuse molecular gas outside the central, CO-bright regions of molecular clouds \citep{wannier93,liszt96,barriault10,allen12,cotten12}, as well as of cold self-absorbing atomic gas in primarily molecular regions \citep{li03,goldsmith05}. Given this, we might expect the OH distribution in the Galactic Plane to be more extended than that of CO. 
However, in the SPLASH pilot region this is not observed.

The channel maps in Fig. \ref{ch1} show $^{12}$CO(J=1--0) from the NANTEN Galactic Plane Survey \citep{mizuno04} together with the four OH lines. The NANTEN data has been smoothed with a Gaussian kernel of FWHM=$15$ arcmin to match the resolution of the SPLASH cubes, and both datasets are binned to a velocity resolution of $3.6$~km~s$^{-1}$, both in order to improve the signal-to-noise, and to allow the inclusion of the full velocity range of interest without excluding data. In general, OH is always seen where there is CO, but there is no indication that it extends outside the boundaries of CO clouds. In fact, the CO is the more extended and readily-detectable of the two, with $\sim70$ per cent of CO-detected voxels having no OH counterpart. While much of this can be attributed to higher signal-to-noise ratios in the CO (such that OH line wings are below the detection threshold where CO line wings are not), the fact remains that many CO lines 
have no detected OH at all. Conversely, OH in the absence of CO makes up only $\sim1$ per cent of total OH detections (excluding masers). Here, we consider the OH detection limit to be $\sim30$ mK, which is the $3\sigma$ sensitivity for data binned to a channel width of $\sim1.8$~km~s$^{-1}$ -- the largest value for which the narrowest diffuse OH spectral line features are  marginally resolved.  

When examining why CO is more widely detected than OH, a key issue is the continuum background level and its effect on detectability. While the noise levels in the SPLASH cubes are not dissimilar to other studies that have detected extended OH envelopes \citep{wannier93,liszt96,barriault10,allen12,cotten12}, particularly when the data are binned over multiple channels, these studies have observed either high latitude or outer Galaxy clouds, where $T_{\mathrm{c}}$ is close to the CMB. In contrast, the continuum emission in the SPLASH pilot region ranges from $\sim8$--45 K, with a mean of $\sim12$ K. 

\begin{figure*}\includegraphics[scale=0.65]{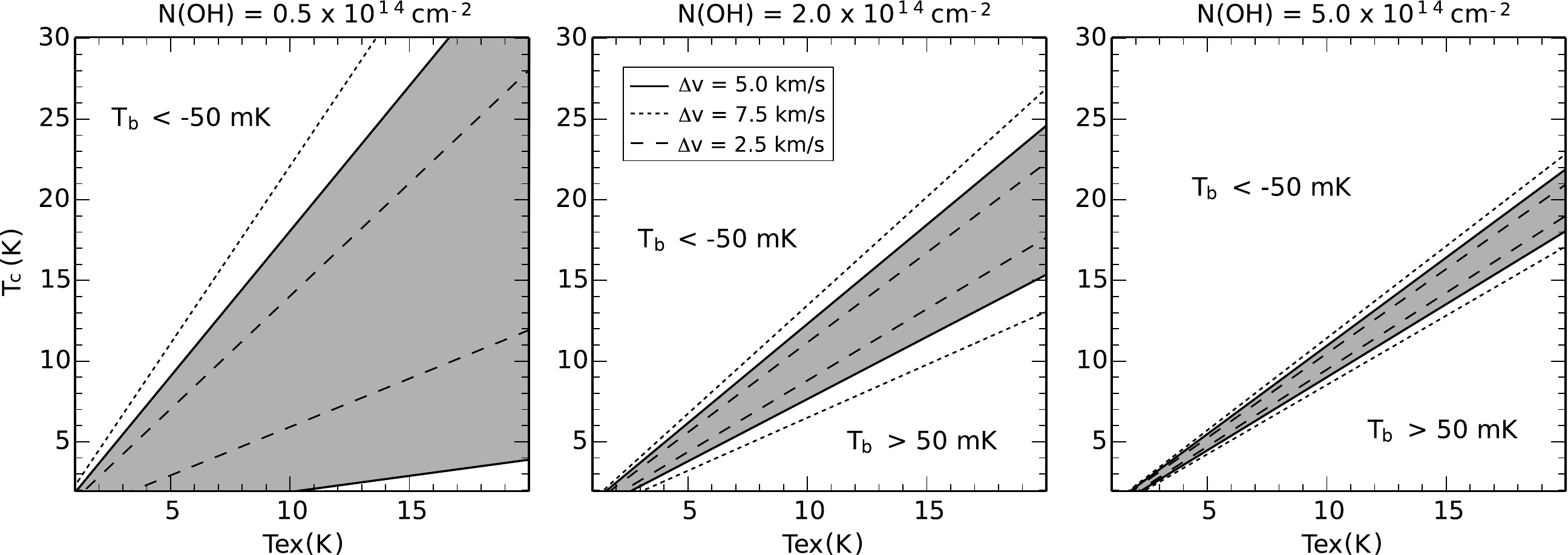}
\caption{Detectability of the 1667-MHz line as a function of total OH column density, $N\mathrm{(OH)}$, line excitation temperature, $T_{\mathrm{ex}}$, and continuum background temperature, $T_{\mathrm{c}}$. Dotted, solid and dashed lines show $\pm50$ mK peak brightness temperature contours for Gaussian line profiles of FWHM of $\Delta v = 2.5$, 5.0 and 7.5~km~s$^{-1}$. This is the approximate $3\sigma$ detection limit for the SPLASH data. The grey shaded area corresponds to $|T_{\mathrm{b}}(1667)| < 50$ mK for the $\Delta v = 5.0$~km~s$^{-1}$ case, illustrating the region of parameter space in which the line is undetectable.}
\label{detectplot}
\end{figure*}

Figure \ref{detectplot} shows 1667-MHz detectability as a function of $T_{\mathrm{ex}}(1667)$ and $T_{\mathrm{c}}$ for OH column densities of 0.5, 2.0 and $5.0\times10^{14}$ cm$^{-2}$, where these values cover a representative range measured in the Galactic Plane and nearby molecular clouds \citep{rieu76,crutcher79,dickey81,colgan89}. The plots are computed from the simple radiative transfer case considered in equation (\ref{radtran}), and the standard relation between optical depth, excitation temperature and the total ground-state OH column density in cm$^{-2}$, 
\begin{equation}
N\mathrm{(OH)}=2.39\times10^{14}~T_{\mathrm{ex}}(1667)~\tau_0(1667)~\Delta v,
\end{equation}
where $\tau_0(1667)$ is the peak optical depth of a Gaussian line profile with FWHM $\Delta v$ (in km~s$^{-1}$). We consider $T_{\mathrm{c}}$ ranging from 3--30 K, and $T_{\mathrm{ex}}(1667)$ limits of 3 and 20 K, where the $T_{\mathrm{ex}}$ values are based on existing direct measurements of both absorbing and emitting gas \citep{dickey81,colgan89,liszt96}. 

In the low-$\tau$ regime in which we are working, the parameter determining detectability for a given column density is $(T_{\mathrm{ex}}-T_{\mathrm{c}})/T_{\mathrm{ex}}$. It can be seen that for the lowest examined column density, main-line OH is undetectable across much of the parameter space, with only high background continuum temperatures guaranteeing detectability (in absorption). This makes detection particularly challenging in the off-Plane regions of the datacube, where column densities are likely on the low-end of the scale, and $T_{\mathrm{c}}$ is not high. For higher column densities the situation improves, and for a moderate $N(\mathrm{OH})$ of $2.0\times10^{14}$ cm$^{-2}$, gas with $|(T_{\mathrm{ex}}-T_{\mathrm{c}})/T_{\mathrm{ex}}| \gtrsim 0.2$ will be detected in the SPLASH cubes. For continuum levels typical of the off-Plane portions of the pilot region ($T_{\mathrm{c}}\sim8$--12 K), this translates to a requirement of $|T_{\mathrm{ex}}-T_{\mathrm{c}}|\gtrsim 2$ K. 

The satellite lines offer an alternative probe, since they are generally out of equilibrium and their excitation temperatures either negative, very positive, or strongly sub-thermal. As discussed in section \ref{results}, around 10 per cent of all diffuse OH detections are seen in the satellite lines alone, demonstrating that small $|T_{\mathrm{ex}}-T_{\mathrm{c}}|$ is indeed responsible for the lack of main-line detection in these cases at least. However, 
the lower transition strength of the satellite lines means that the factor $|(T_{\mathrm{ex}}-T_{\mathrm{c}})/T_{\mathrm{ex}}|$ must be approximately 9 times larger than for the 1667-MHz line in order to produce radiation of the same intensity. 
For the moderate column density case this sets an approximate detection limit of $|(T_{\mathrm{ex}}-T_{\mathrm{c}})/T_{\mathrm{ex}}| \gtrsim 2$. While the distribution of OH satellite-line excitation temperatures has not yet been determined on large scales in the Galaxy, it is likely that satellite-only detections will be limited either to regions of moderate column density, or to where the population inversion (and anti-inversion) is particularly strong.

More stringent quantitative limits on the OH column density await the results of deeper pointed observations, as well as an in-depth assessment of the satellite-line excitation state across the wider survey region. 
However, from a practical perspective we may conclude that studies of the transition-state molecular ISM using OH are best conducted either towards regions of either significantly high or significantly low continuum brightness. The moderate continuum brightness temperatures seen in much of the SPLASH pilot cube are not conducive to such studies, and are likely in part responsible for the lack of OH detection from molecular cloud envelopes.

\subsection{Maser sources in the pilot region}
\label{masers}

Throughout the pilot survey region, we detect masers in all four OH transitions. 
As described in Section \ref{ohbasics}, we identify as a maser or maser candidate any source that fulfils the following criteria: 

\begin{itemize}
\item The emission appears unresolved in the Parkes beam,
\item The spectral profile shows narrow features,
\item The pattern of matching or symmetrical features typical of diffuse OH is not seen in the other three lines.
\end{itemize}

\noindent In this context, the term `maser candidate' primarily applies to those objects with flux densities 
close to the SPLASH detection limit, since it is possible that such features may be false detections, and/or that corresponding profiles in the other 3 lines are present just below the detection threshold. 

The approximate locations of the maser sources are shown in Fig. \ref{mom-4tiles}. We do not provide positions in this work, since the Parkes beam does not allow precise positional determination. Accurate positions for these masers are currently being obtained with follow-up work using the ATCA and will be reported in a future publication. 

\begin{figure}
\begin{center}
\includegraphics[scale=0.6]{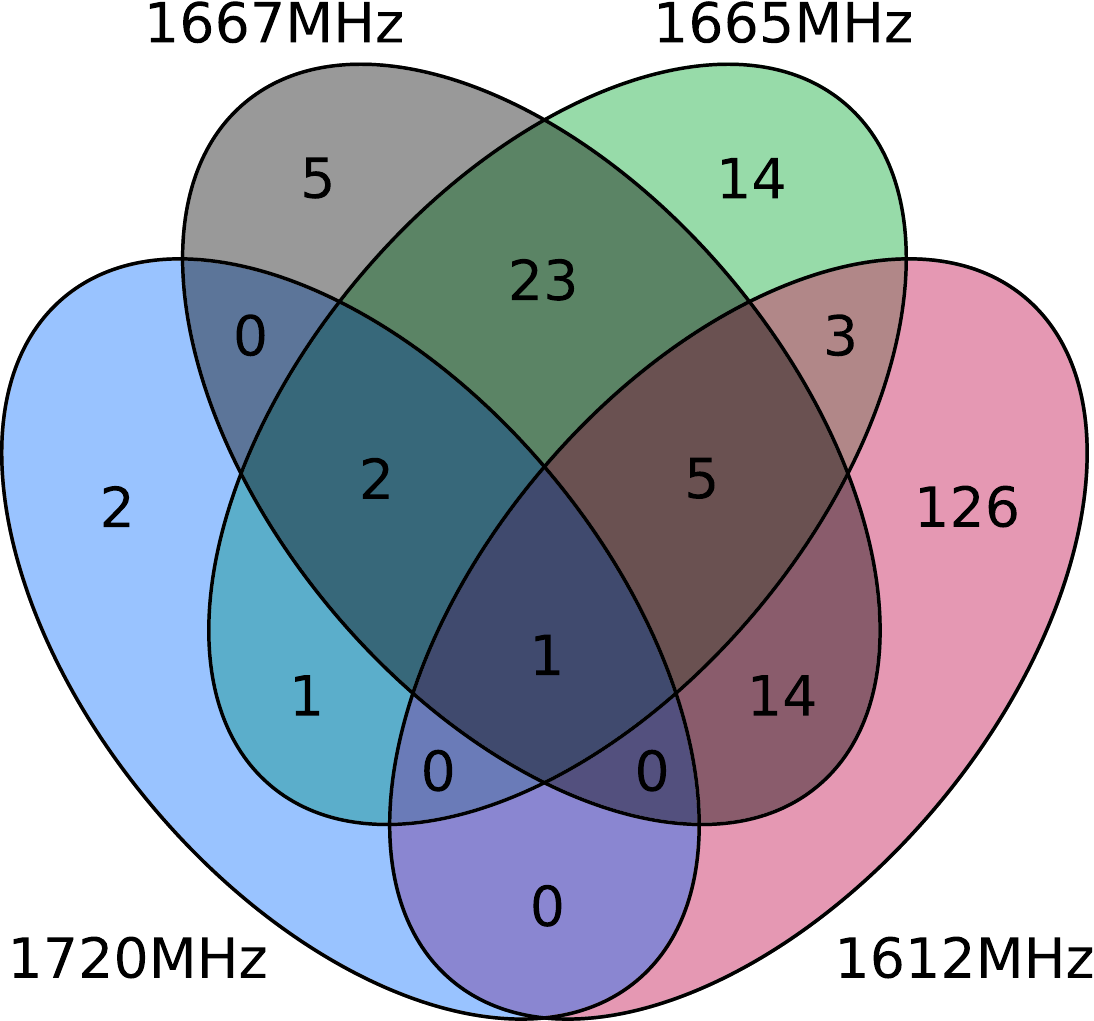}
\caption{Venn diagram showing the overlap in occurrence of masers and maser candidates in the four OH lines.}
\label{vennpic}
\end{center}
\end{figure}

We find a total of 196 sites which contain maser sources. Of these, 149 show emission in the 1612-MHz line, 49 in the 1665-MHz line, 50 in the 1667-MHz line and 6 in the 1720-MHz line. Figure \ref{vennpic} shows a Venn diagram of the overlap 
between the four transitions. Here, two masers are considered to be coincident if their peak positions occupy either the same or adjacent pixels, and their velocity ranges overlap. We note that given the large Parkes beam and the relatively loose requirement for agreement in velocity, we cannot be certain that such associations are genuine -- either in the sense of emission in both lines arising from physically associated parcels of gas, or 
in the sense of being excited by the same powering source. 

Main-line masers show the largest overlap; of the 1665- and 1667-MHz masers, 31 are present in both transitions. For main-line masers in star formation regions, \citet{caswell98} combined the results of \citet{caswell83,caswell87} to find that 90 per cent of 1665-MHz OH masers have a 1667-MHz counterpart. SPLASH finds a somewhat lower fractional association rate (63 per cent). This deficit of associations likely arises in part from 
contamination from evolved star masers that have been included in our association statistics, but this may not explain the entirety of the discrepancy. \citet{caswell98} noted that the median flux density ratio of 1665- to 1667-MHz sources is 3. 
This flux density ratio has considerable scatter and may exhibit a dependence on the luminosity of the 1665-MHz maser emission, similar to 6.7-GHz methanol to OH maser ratios \citep{caswell97} and 6.7-GHz methanol to 12.2-GHz methanol maser ratios \citep{breen11b}. If this were the case, it is conceivable that the weaker sources we detect in general have a larger flux density ratio, pushing the 1667-MHz counterpart below our detection threshold in some cases. A further possibility is that our much more sensitive observations have uncovered a population of sources with properties different to those catalogued by \citet{caswell83,caswell87}. Each of these scenarios with be investigated in future publications using the full SPLASH OH maser data, as well as with high-resolution follow-up observations. 

The majority of the 19 1667-MHz OH masers that are not associated with a 1665-MHz source are associated with 1612 MHz maser emission (14 of 19),
and in general consist of weak emission at the velocity of one or both of the peaks in the 1612-MHz double-horned profile. Four of the six 1720-MHz OH maser detections are associated with 1665-MHz emission, with a mixture of additional accompanying transitions. One of the solitary 1720-MHz OH masers, 334.10+0.40, is a new detection, and may be of the variety noted by
\citet{caswell04} as possibly associated with star formation despite the absence of main-line emission. 


The pilot region has previously been surveyed for OH main-line maser emission 
by untargetted mosaicing observations with the ATCA \citep{caswell98}, allowing us to compare our detection statistics. 
This survey covered Galactic latitudes of approximately $-1^\circ < l < +1^\circ$. Its frequency coverage allowed for the detection of the 1665-MHz line at Galactic velocities, and the 1667-MHz line only at velocities less negative than $-80$~km~s$^{-1}$. For this reason we compare only the 1665-MHz line here. Where the \citeauthor{caswell98} and SPLASH survey regions overlap, \citeauthor{caswell98} detected 37 1665-MHz OH maser sites, all but one of which were also detected by SPLASH. These 37 sites correspond to 29 distinct positions in the Parkes data, since in some cases more than one site falls within within the same Parkes beam. In addition to this, we detect a further 12 1665-MHz maser sources in our data that were not detected by \citeauthor{caswell98} and a further 11 outside his survey region. i.e. in the region which has previously been surveyed we find approximately 30 per cent of our masers and maser candidates are new detections. If follow-up observations confirm that the majority of these new candidates are indeed masers, then the complete sampling and increased sensitivity of SPLASH compared to previous work will have produced a substantial increase in the known population for this transition. In regions which have not been subject to untargetted surveys the fraction of newly detected masers is expected to be greater.

\begin{figure}
\begin{center}
\includegraphics[scale=0.4]{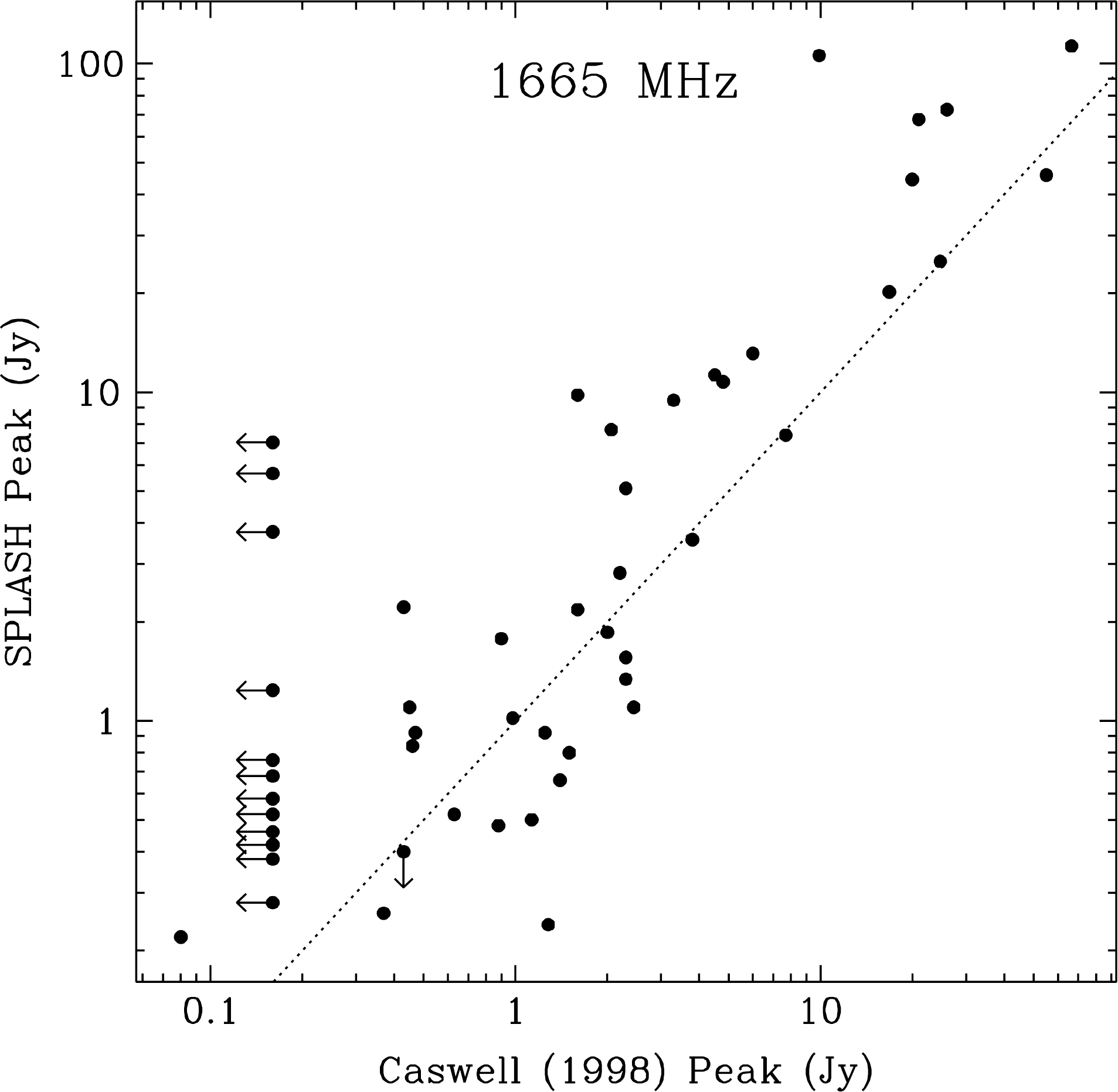}
\caption{Distribution of peak flux densities of 1665-MHz maser sources as detected by SPLASH and \citet{caswell98}. The dotted line represents equality between the two flux density measurements. Non-detections \citeauthor{caswell98}'s data are plotted at his $5\sigma$ detection limit of 160~mJy, with arrows pointing left. The single source reported by Caswell that is not detected in the SPLASH data is plotted at a $5\sigma$ detection limit of 400~mJy (higher than most of the survey region due to the presence of bright continuum at that position) with an arrow pointing downwards.}
\label{smplot2}
\end{center}
\end{figure}

In Fig. \ref{smplot2} we show the distribution of peak flux densities for our and Caswell's data. For those masers detected in both observations, there is a reasonably good one-to-one correspondence in the peak flux densities, with some scatter. 
This suggests that the 1665-MHz masers typically vary by a factor of a few over the timescale between the observations for the two datasets (16-20 years). However, three non-detections in the Caswell data have substantial peak flux densities in our data (between 3.8 and 7.0\,Jy). The most likely explanation for this is that these three objects have exhibited particularly strong flux density variations. We note the appearance of a significant asymmetry in the sources detected on a single epoch: 12 sources detected by SPLASH and not Caswell, compared to 1 detected by Caswell and not by SPLASH. The larger number of sources detected only in SPLASH may be largely accounted for by the slightly less sensitive Caswell data, especially given the non-uniform sampling of the Galactic plane (8 of these 12 sources have flux densities significantly less than 1~Jy).   

\begin{figure}
\begin{center}
\includegraphics[scale=0.4]{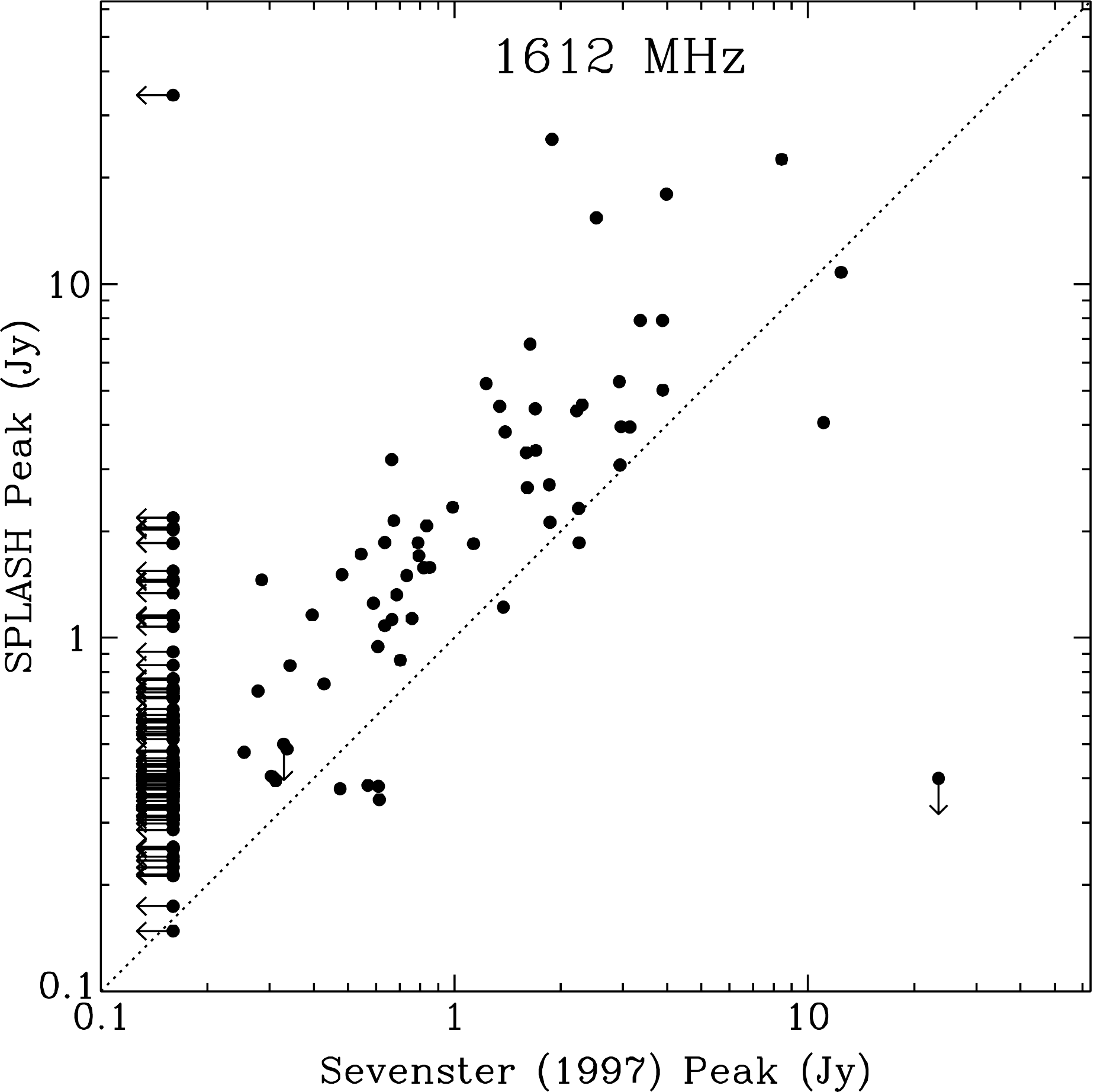}
\caption{Distribution of peak flux densities of 1612-MHz maser sources as detected by SPLASH and \citet{sevenster97}. The dotted line represents equality between the two flux density measurements. Non-detections in \citeauthor{sevenster97}'s data are shown at their $5\sigma$ detection limit of 175~mJy, with arrows pointing left. Non-detections in the SPLASH data are plotted at the $5\sigma$ detection limits appropriate for their locations, with arrows pointing downwards.}
\label{smplot3}
\end{center}
\end{figure}

\citet{sevenster97} surveyed the entire SPLASH pilot region for masers in the 1612-MHz line. They identified 65 masers within this region, all but two of which are detected in our observations. Here we require the peak position of the maser in the Sevenster catalogue to fall either on or adjacent to the peak pixel position of the SPLASH source, and that the velocities ranges of the two sources overlap. Of the 61 detected sources, 59 appear as distinct sites within the Parkes beam. In addition to this, we detect 91 maser sources that were not detected in Sevenster's work. A comparison of the distribution of peak flux densities between the two datasets (Fig. \ref{smplot3}) shows a scatter that is partly due to intrinsic variability. 
Generally, the points fall to the left of the dotted line of equality in the figure. This is due to a lower reported peak flux density in the \citeauthor{sevenster97} data, arising 
from a spectral channel width that is considerably larger than that of the present work (1.46~km~s$^{-1}$ compared to 0.18~km~s$^{-1}$). This tends to reduce the measured peak flux density when emission features have line widths that are comparable to (or smaller than) the channel width. \citeauthor{sevenster97} noted this and compared their results to higher spectral resolution observations of a subset of their sample, finding that their peak flux densities were reduced by factors of between 1.3 and 1.8 compared to the higher spectral resolution data. We find that the \citet{sevenster97} peak flux densities are, on average, 1.75 times less than the SPLASH peak flux densities, consistent with their analysis. The scatter in Fig. \ref{smplot3} is also increased by this effect.

Overall, approximately 60 per cent of the 1612-MHz masers and maser candidates in the SPLASH pilot region are new detections. This large number is likely a result not only of improved sensitivity, but also because of the differences in velocity resolution and detection criteria in the two surveys. 
The per-channel rms sensitivity of \citeauthor{sevenster97}'s catalogue is $\sim35$~mJy in 90 per cent of their fields.
When binned to the same channel width as Sevenster et al.'s study, SPLASH is more sensitive by a factor of $\sim$1.5 (the mean unbinned per-channel rms sensitivity is $\sim$65 mJy). Additionally, \citet{sevenster97b} require detections in three (not necessarily consecutive) velocity channels for a source to qualify for inclusion in their catalogue (or four channels for weak emission), meaning sources with intrinsic linewidths less than 4.4~km~s$^{-1}$ (5.8~km~s$^{-1}$ in the weak case) are more easily rejected.
We find that SPLASH sources that were not detected by \citet{sevenster97} are typically sources that are either narrow, relatively weak ($\lesssim 1$~Jy in SPLASH) and/or have only a single spectral component. The absence of the strongest 1612-MHz maser (33.9~Jy) detected in the SPLASH observations, 343.119-0.067 \citep{caswell99}, may also be attributable to its narrow linewidth of (FWHM) $\Delta v\sim0.7$~km~s$^{-1}$.

It is also likely that intrinsic variations in the peak flux densities are responsible for some differences in source detection in the two catalogues. In particular, we note that the strongest 1612-MHz maser detected by \citeauthor{sevenster97} in the pilot region, 338.925+0.557, was not detected by SPLASH. 
This source coincides with maser emission from methanol \citep{caswell11} and OH at 1665-, 1667- and 1720-MHz \citep{caswell99,caswell04} at a site of star formation, with all transitions displaying strong variability. This strong intrinsic variability is a likely explanation for its absence from SPLASH.  


\section{Summary and conclusions}

We have presented the first results from SPLASH -- a sensitive, fully-sampled and unbiased survey of all four 18-cm ground-state transitions of OH in the Southern Galactic Plane and Galactic Centre region. The pilot region ($334^{\circ} < l < 344^{\circ}$, $|b| < 2^{\circ}$) presented here covers 40 square degrees of the inner Galaxy, including several prominent spiral arms and the 3 kpc Arm tangent. The 1612-, 1665-, 1667- and 1720-MHz spectral line cubes (optimised for weak, extended emission) have a per-channel rms sensitivity of $\sim16$ mK in a (binned) 0.7~km~s$^{-1}$ channel and an effective resolution of $\sim15.5$ arcmin. The sensitivity of the survey to unresolved maser sources is $\sim65$ mJy in a 0.18~km~s$^{-1}$ channel. SPLASH also provides complementary continuum maps at 1612, 1666 and 1720 MHz, with an estimated uncertainty of $\lesssim1.0$ K. 

Diffuse OH is detected in all four transitions throughout the SPLASH pilot region, tracing prominent Galactic structures. The lines show a characteristic signature: main lines with $|T_{\mathrm{b}}(1667)| \gtrsim |T_{\mathrm{b}}(1665)|$ (usually seen in absorption), and a symmetrical pattern of emission and absorption in the satellite lines. In approximately 10 per cent of detected voxels, signal is seen in the satellite lines alone, highlighting their largely-untapped potential as tracers of the diffuse molecular ISM. In such cases the lack of main-line detection must arise from main-line excitation temperatures that are within a couple of degrees of the diffuse continuum background. 

Indeed, the level of the continuum background is an important determinant of OH line detectability. While other studies at similar sensitivities have observed OH envelopes extending beyond the central CO-bright regions of molecular cloud complexes, we do not observe this phenomenon in the present work. 
We argue that small values of $|T_{\mathrm{ex}}-T_{\mathrm{c}}|$ are at least in part responsible for this lack of detection, particularly in the off-Plane regions of the cube, where column densities are likely to be low, and excitation temperatures are likely close to the continuum brightness temperature ($T_{\mathrm{c}}\sim8$--12 K). 
While the satellite lines (which are always anomalously excited) offer an alternative probe, their smaller transition strengths present a different obstacle to detection. Quantitative limits on the OH column density await the results of deeper pointed observations, as well as an in-depth assessment of excitation patterns across the wider
survey region. Nevertheless, it is evident that studies of the transition-state ISM are best conducted toward regions of either significantly high or significantly low background continuum emission. 

From simultaneous examination of all four transitions, we conclude that diffuse OH optical depths (averaged over the Parkes beam) are small throughout the pilot region -- an assumption which is likely to remain valid throughout the Milky Way. 
We also find that the assumption that the main lines are in LTE produces unphysical results, implying that `anomalous' main-line excitation is the norm throughout this section of the Galactic Disk. These results highlight the power of matching observations of the full set of ground state lines, and provide an important basis for analysis of the full survey.

Finally, preliminary analysis of the population of ground-state OH masers in the region suggests that the complete sampling and high sensitivity of SPLASH will result in the detection of a substantial population of hitherto unknown maser sources. 
Approximately 50 per cent of the OH masers and maser candidates identified in the pilot region are new detections, with the majority of these in the 1612-MHz line. Scatter in flux density between the present work and previous surveys provides some constraints on variability, which in some cases includes the absence of strong masers that were previously observed, or the detection of strong sources not present in previous work. Follow-up observations of the SPLASH source sample with the ATCA are underway.


\section*{Acknowledgments}

We wish to thank Stacy Mader, John Reynolds, Brett Preisig, Mal Smith and the entire Parkes telescope staff for their support and advice during the substantial hours of SPLASH project observing. We would also like to thank the referee, Paul Goldsmith, for his prompt and helpful report. The Parkes Telescope is part of the Australia Telescope which is funded by the Commonwealth of Australia for operation as a National Facility managed by CSIRO. The NANTEN project was based on a mutual agreement between Nagoya University and the Carnegie Institute of Washington, and its operation was made possible thanks to contributions from many companies and members of the Japanese public. SLB acknowledges support from the Australian Research Council. JFG acknowledges support from grant AYA2011-30228-C03-01 of MINECO (which includes FEDER funds). NL's postdoctoral fellowship is supported by a CONICYT/FONDECYT postdoctorado, under project no. 3130540. 

\footnotesize{
\bibliography{splashbib}}

\appendix
\section{Channel maps}

\begin{figure*}\includegraphics[scale=0.93]{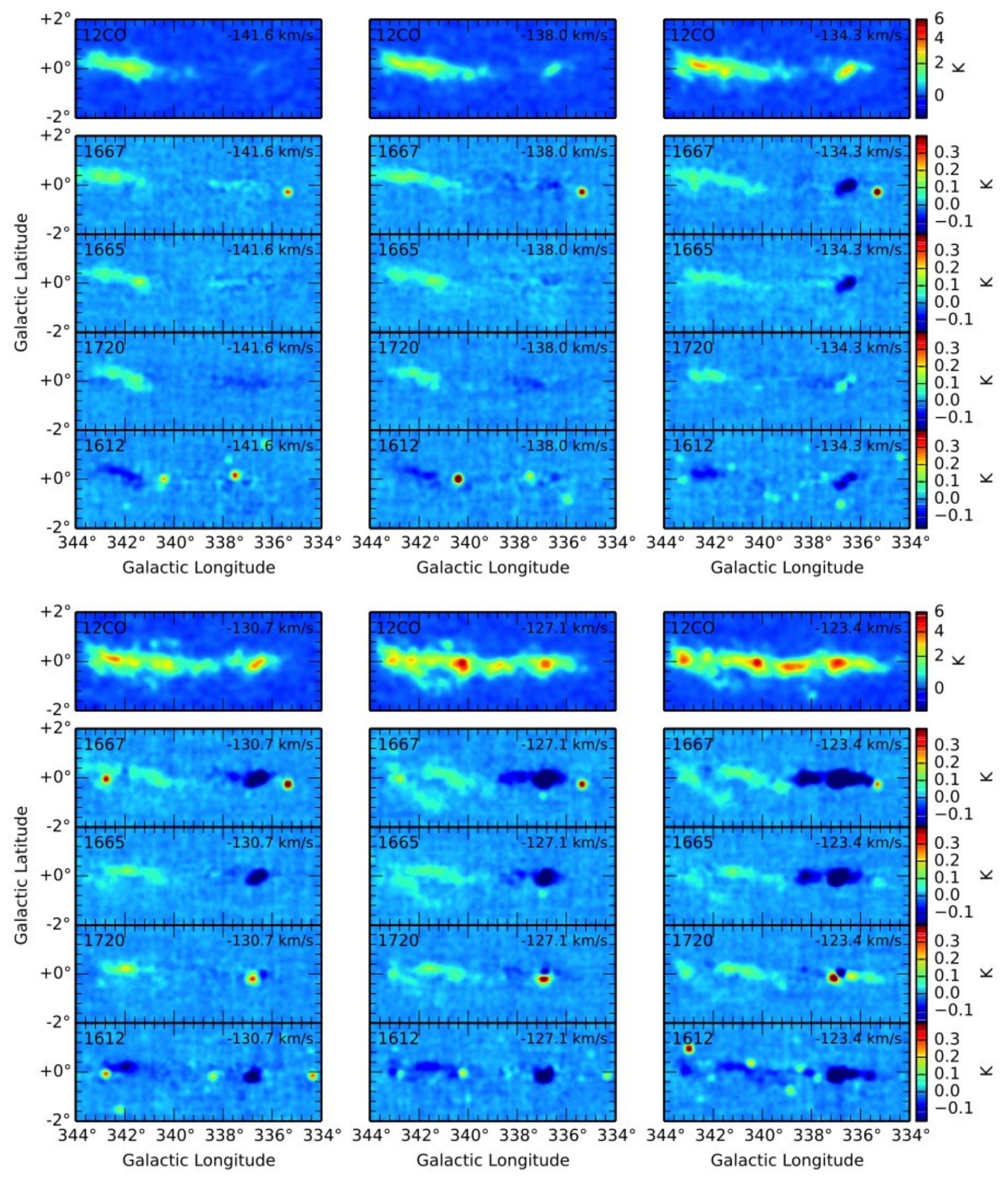}
\caption{Velocity channel maps ($v_{LSR}$) of all four ground state OH lines. Each panel shows the mean brightness temperature in a 3.7~km~s$^{-1}$ interval. $^{12}$CO(J=1--0) data from the NANTEN Galactic Plane Survey is shown for comparison, smoothed to a spatial resolution of $15.5$ arcmin.}
\label{ch1}
\end{figure*}

\addtocounter{figure}{-1}
\begin{figure*}\includegraphics[scale=0.93]{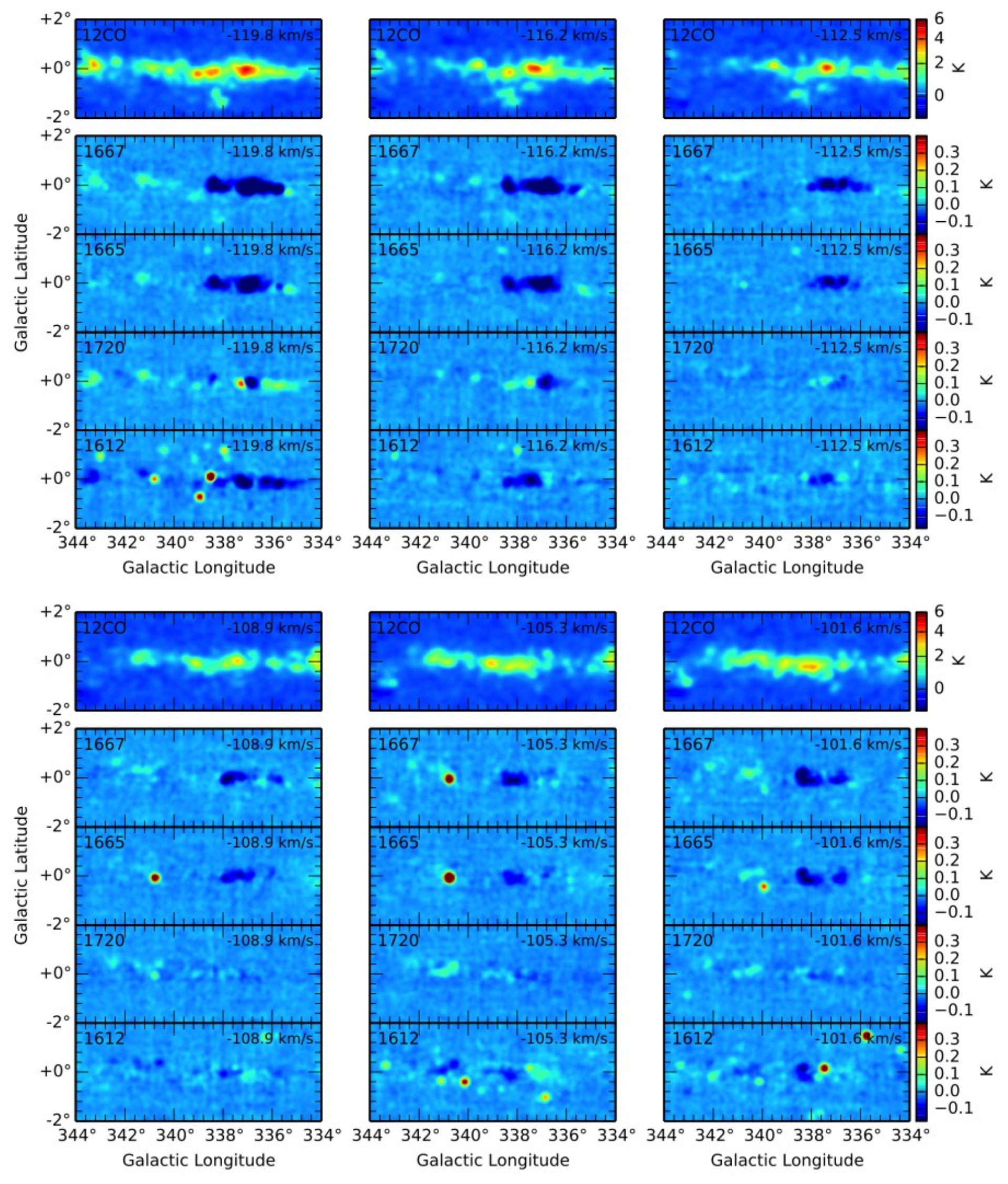}
\caption{ -- cont.}
\label{ch2}
\end{figure*}

\addtocounter{figure}{-1}
\begin{figure*}\includegraphics[scale=0.93]{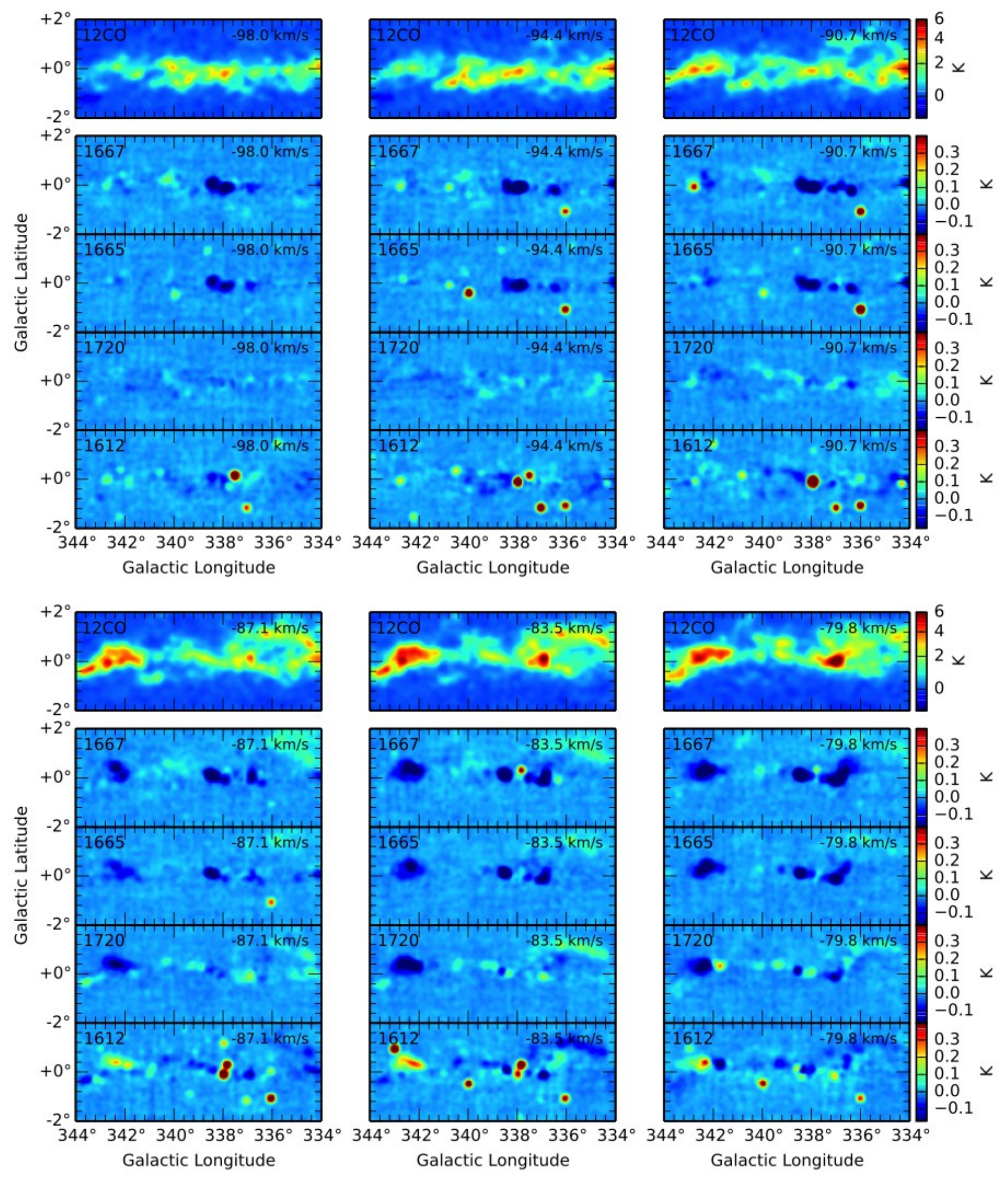}
\caption{ -- cont.}
\label{ch3}
\end{figure*}

\addtocounter{figure}{-1}
\begin{figure*}\includegraphics[scale=0.93]{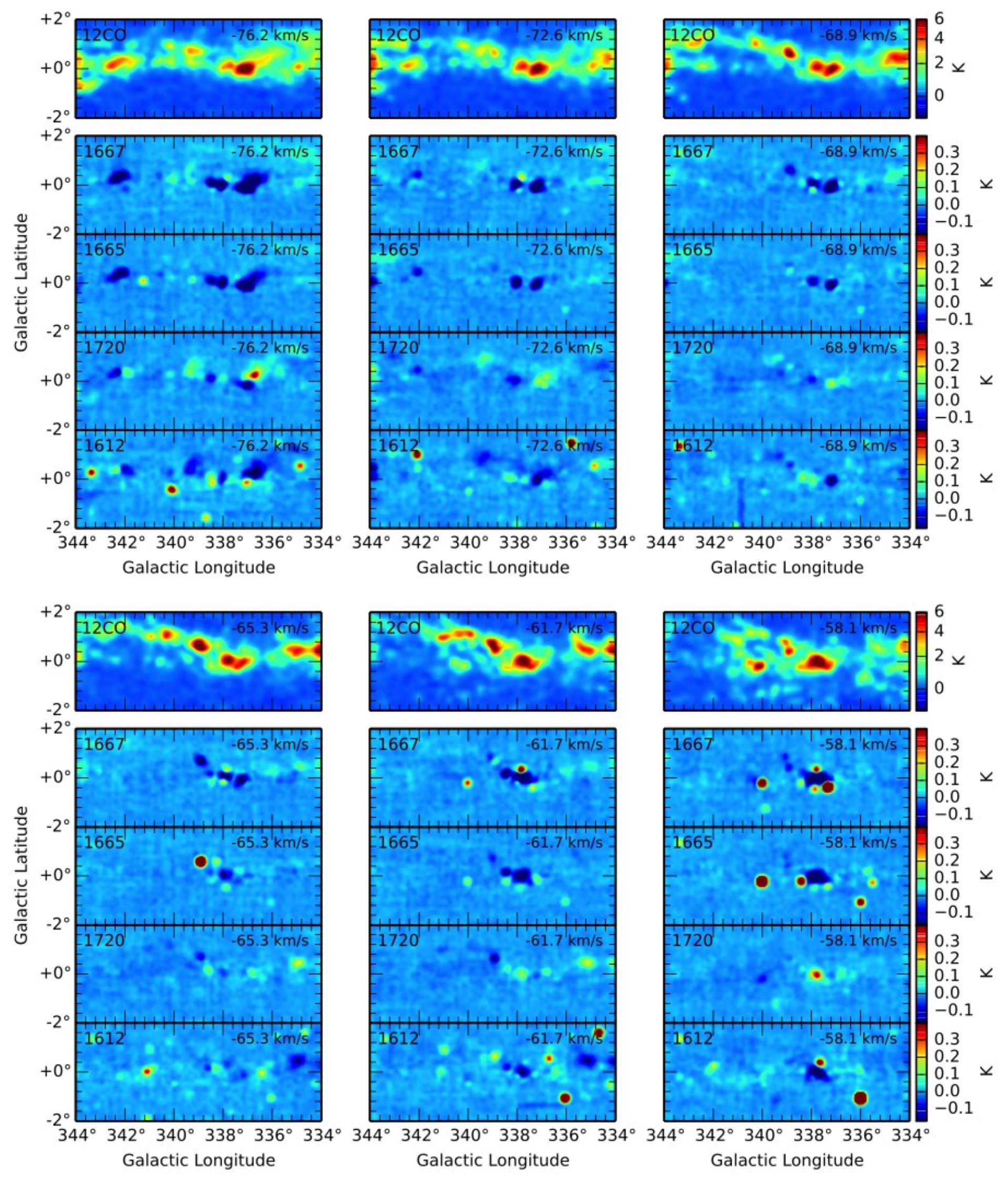}
\caption{ -- cont.}
\label{ch4}
\end{figure*}

\addtocounter{figure}{-1}
\begin{figure*}\includegraphics[scale=0.93]{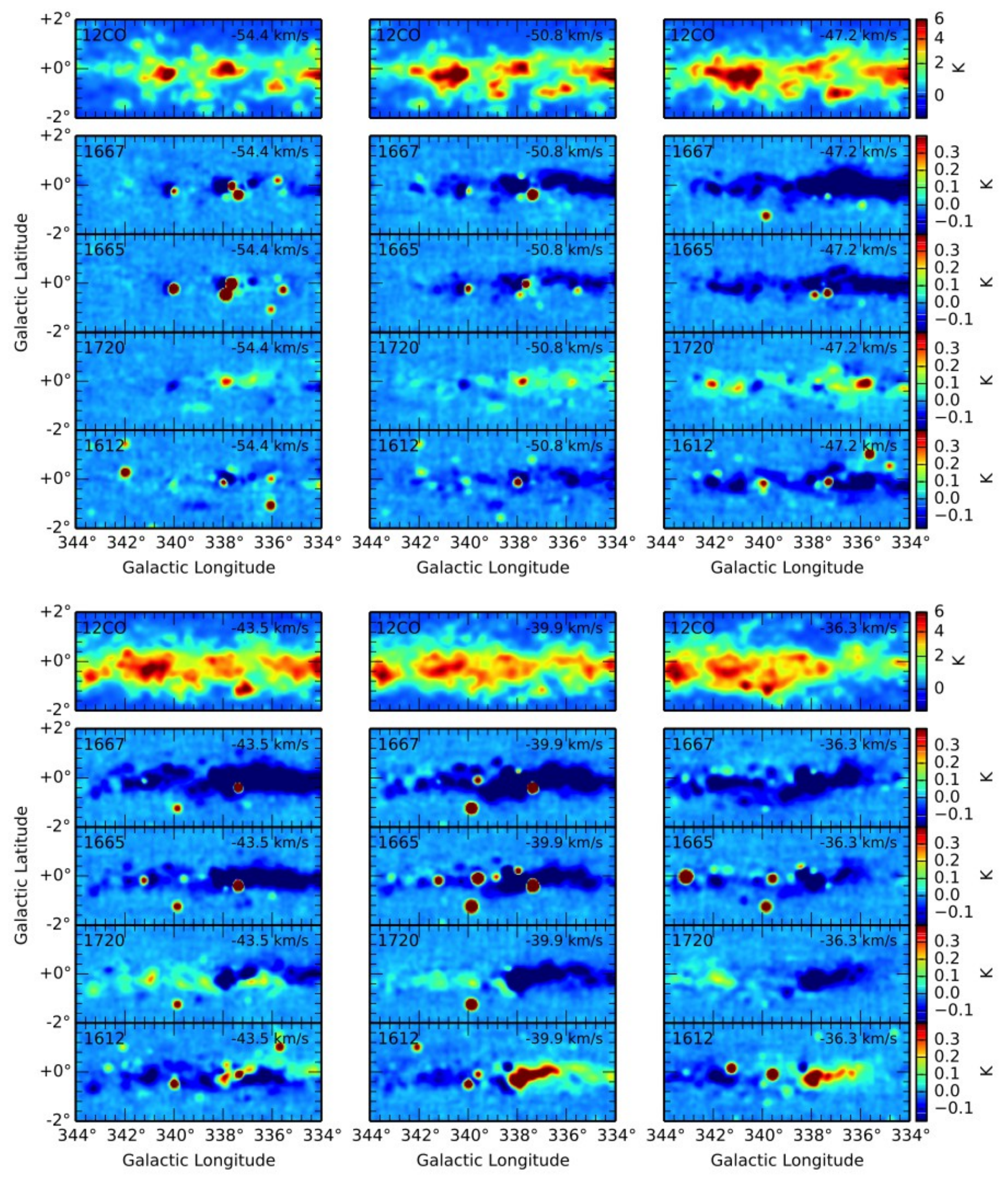}
\caption{ -- cont.}
\label{ch5}
\end{figure*}

\addtocounter{figure}{-1}
\begin{figure*}\includegraphics[scale=0.93]{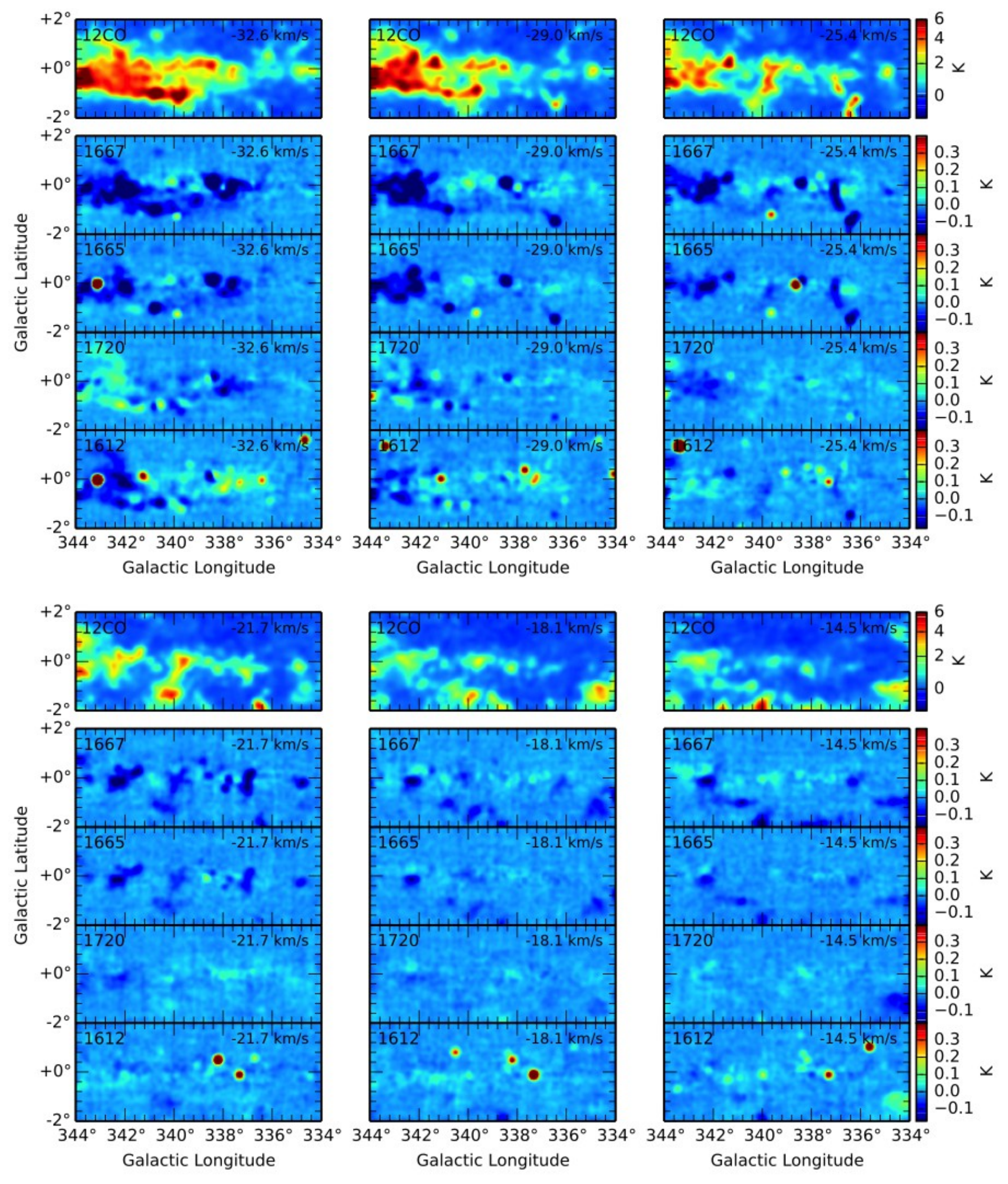}
\caption{ -- cont.}
\label{ch6}
\end{figure*}

\addtocounter{figure}{-1}
\begin{figure*}\includegraphics[scale=0.93]{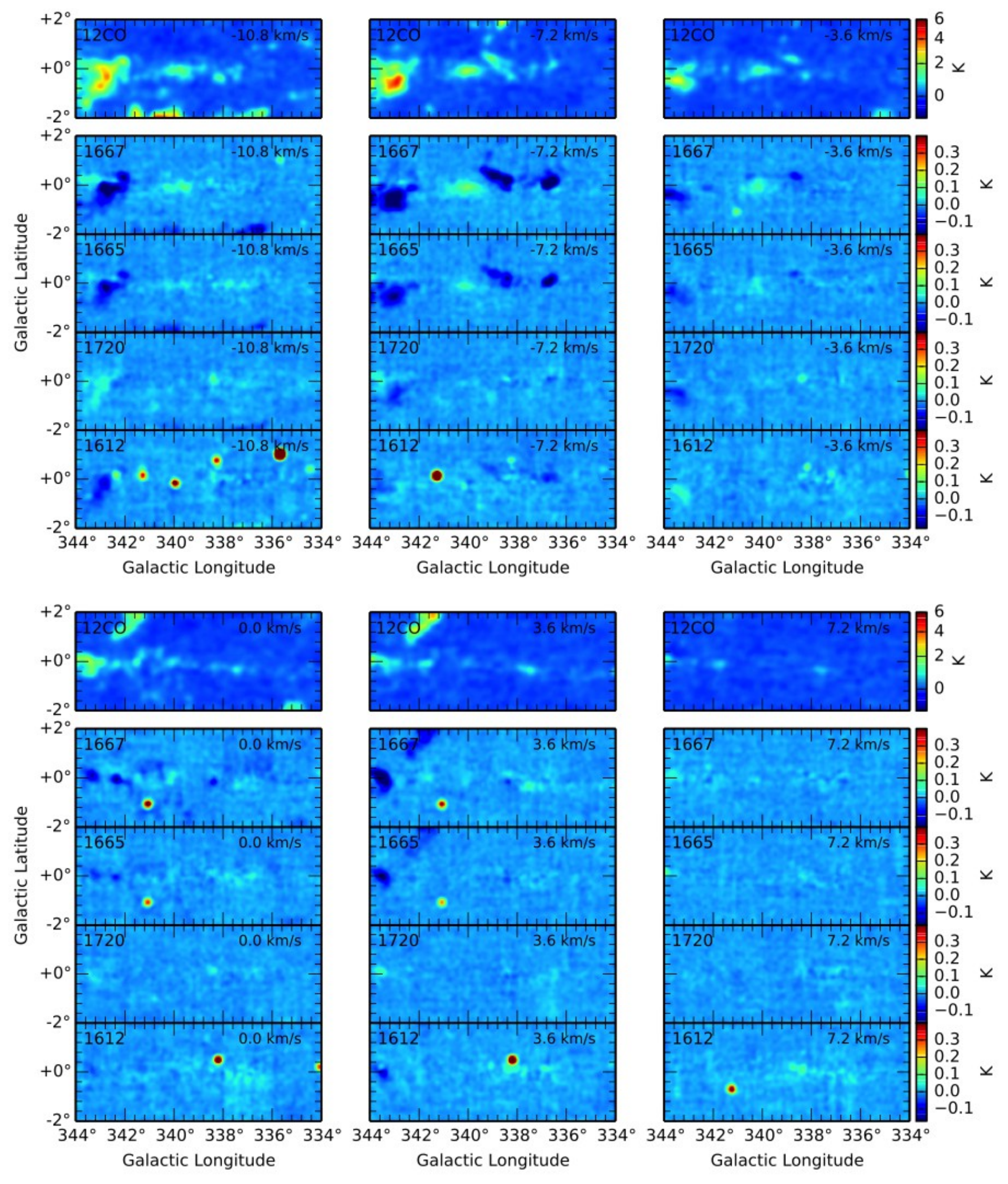}
\caption{ -- cont.}
\label{ch7}
\end{figure*}

\end{document}